\documentclass[onecolumn,a4paper]{revtex4}

\usepackage{graphicx}
\usepackage{epstopdf}
\usepackage{graphicx}
\usepackage{bm}
\usepackage[latin1]{inputenc}
\usepackage{amsmath}
\usepackage{amsfonts}
\usepackage{subfigure}
\usepackage{amssymb}
\usepackage{color}
\usepackage{adjustbox}
\usepackage{lipsum}
\usepackage{mathrsfs}
\usepackage{amsfonts}
\RequirePackage[colorlinks,citecolor=blue,urlcolor=blue,linkcolor=magenta]{hyperref}

\begin{document}

\title{Scalar-Fluid interacting dark energy:\\ cosmological dynamics beyond the exponential potential}

\smallskip

\author{\bf~ Jibitesh~Dutta$^{1,2}$\footnote{jdutta29@gmail.com, jdutta@associates.iucaa.in}, Wompherdeiki Khyllep$^3$\footnote{sjwomkhyllep@gmail.com}, Nicola Tamanini$^4$\footnote{nicola.tamanini@cea.fr} }

\smallskip

\affiliation{$^{1}$Mathematics Division, Department of Basic Sciences and Social Sciences, North Eastern Hill University,
NEHU Campus, Shillong, Meghalaya 793022, India}

\affiliation{$^{2}$ Inter University Centre for Astronomy and Astrophysics, Pune 411 007, India}

\affiliation{$^{3}$ Department of Mathematics, St. Anthony's College, Shillong, Meghalaya 793001, India}

\affiliation{$^4$Institut de Physique Th\'eorique, CEA-Saclay,
CNRS UMR 3681, Universit\'e Paris-Saclay, F-91191 Gif-sur-Yvette, France }

\date{\today}

\begin{abstract}
 We extend the dynamical systems analysis of Scalar-Fluid interacting dark energy models performed in C.~G.~Boehmer et al Phys.\ Rev.\ D {\bf 91}, 123002 (2015), by considering scalar field potentials beyond the exponential type.
 The properties and stability of critical points are examined using a combination of linear analysis, computational methods and advanced mathematical techniques, such as centre manifold theory.
 We show that the interesting results obtained with an exponential potential can generally be recovered also for more complicated scalar field potentials.
 In particular, employing power-law and hyperbolic potentials as examples, we find late time accelerated attractors, transitions from dark matter to dark energy domination with specific distinguishing features, and accelerated scaling solutions capable of solving the cosmic coincidence problem.
\end{abstract}

\maketitle


\section{Introduction}\label{sec:Intro}

As by now confirmed by precise cosmological observations \cite{Riess:1998cb,Perlmutter:1998np,Betoule:2014frx,Ade:2013zuv,Ade:2015xua}, our Universe is presently undergoing through a period of accelerated expansion.
In the standard cosmological picture, this phenomenon can be explained by an exotic repulsive cosmic fluid known as dark energy (DE), whose fundamental nature is still unclear.
The easiest theoretical model for DE, which well fits the present cosmological observations \cite{Betoule:2014frx,Ade:2013zuv,Ade:2015xua}, is the so called $\Lambda$CDM model, resulting by a simple addition of a positive cosmological constant $\Lambda$ to the Einstein field equations.
This model accounts for both DE, through $\Lambda$, and dark matter (DM), the other invisible component needed to match the astronomical data, through a pressure-less fluid which does not interact with electromagnetic radiation.
Unfortunately, although well in agreement with observations, $\Lambda$CDM is plagued by unsolved theoretical issues, such as the cosmological constant problem \cite{Weinberg:1988cp,Martin:2012bt} and the cosmic coincidence problem \cite{Steinhardt:1999nw}.

In order to alleviate these problems, a dynamical scalar field, which is capable to reproduce the properties of a cosmological constant at late times, has been proposed as an alternative explanation to the present cosmological acceleration (see \cite{Copeland:2006wr,Tsujikawa:2013fta} for reviews).
Models based on scalar field theories are enough complex to produce a non trivial cosmological dynamics and nonetheless sufficiently simple to handle.
They are collectively known under the name {\it quintessence} and can be well motivated by the lower energy limit of some well known high energy theories like string theory. Moreover in a cosmological context, apart from describing DE, scalar fields are also used to describe inflation \cite{Liddle:2000cg}, DM \cite{Magana:2012ph} and also unified dark sector models \cite{Bertacca:2010ct}.

Once one assumes DE to be a dynamical entity, in contrast with the time-independent cosmological constant, nothing prevents a possible interaction between the two dark sector components, namely between DE and DM.
One of the advantages of considering a dark sector coupling is the existence of late time accelerated scaling attractors, which normally cannot be obtained without an interaction, and can in principle represent a possible solution to the cosmic coincidence problem \cite{Amendola:1999er,Zimdahl:2001ar,Mangano:2002gg,Chimento:2003iea}.
Unfortunately, given the absence of a fundamental satisfying description of both DM and DE, no one knows how to theoretically implement such a coupling, and all the models proposed so far rely
on some simple phenomenological approaches (see \cite{Wang:2016lxa} for a recent review), which in general might give rise to complications at the cosmological perturbation level \cite{Valiviita:2008iv,Skordis:2015yra}.

Recently, a new phenomenological approach for interacting DE theories, which uses a well posed variational method and thus is completely well defined also at the fully covariant level, has been introduced in \cite{Boehmer:2015kta,Boehmer:2015sha} (see \cite{Pourtsidou:2013nha,Tamanini:2015iia} for similar ideas).
In this theory, DM is characterized by a perfect fluid and integrated into a variational principle using Brown's Lagrangian formulation of relativistic fluids \cite{Brown:1992kc}.
The general class of theories defined in this way has been called {\it Scalar-Fluid theories} \cite{Koivisto:2015qua,Boehmer:2015ina}, and besides interacting DE it has also been applied to build models of screened scalar fields \cite{Brax:2015fcf,Tamanini:2016klr}.
Unlike previous interacting DE proposals, e.g.~\cite{Boehmer:2008av,Boehmer:2009tk,Dutta:2016dnt,Wang:2016lxa}, in this approach the interaction is introduced directly in a suitably defined action, whereby the conservation equation is automatically satisfied.
Therefore, this type of coupling has the advantage over the usual phenomenological interactions of being consistently constructed at the Lagrangian level and thus of being well motivated by an underlining theoretical framework, even though not a fundamental one.

In this paper, we extend the cosmological dynamical systems analysis of Scalar-Fluid interacting DE models performed in \cite{Boehmer:2015kta}.
Dynamical system techniques are a useful tool to study the asymptotic behaviour and to determine the complete dynamics of a cosmological model.
For an introduction to the applications of dynamical system in cosmology we refer the reader to \cite{Garcia-Salcedo:2015ora,Boehmer:2014vea} (see also \cite{WainwrightEllis,Coley:2003mj}).
These techniques have been largely applied to several cosmological models.
For some recent studies we refer the reader to the following works: quintessence field \cite{Roy:2014yta,Paliathanasis:2015gga,Alho:2015ila,Landim:2016gpz}, $k$-essence \cite{Tamanini:2014mpa,Dutta:2016bbs}, Brans-Dicke theory \cite{Cid:2015pja,Quiros:2015bfa,Garcia-Salcedo:2015naa}, $f(R)$ gravity \cite{Alho:2016gzi}, hybrid metric-Palatini gravity \cite{Tamanini:2013ltp,Carloni:2015bua}, $f(T)$ theory \cite{Cai:2015emx}, chameleon theories \cite{Roy:2014hsa,Tamanini:2016klr}, holographic DE \cite{Banerjee:2015kva}, braneworld theories \cite{Escobar:2013js,Dutta:2015jaq}, interacting DE \cite{Dutta:2016dnt,Landim:2015poa,Landim:2015uda}.

In \cite{Boehmer:2015kta}, the scalar field potential is taken to be of the exponential kind, leading to three dimensional dynamical systems whose dynamical evolution is rather simple to analyse.
The aim of this paper is to investigate the cosmological dynamics of the same Scalar-Fluid interacting quintessence models considered in \cite{Boehmer:2015kta} for a class of scalar fields whose self interacting potential $V(\phi)$ is left arbitrary.
In order to accomplish our scopes, we will introduce a parameter $\Gamma\left(=V\frac{d^2V}{d \phi^2}\left(\frac{d V}{d \phi}\right)^{-2}\right)$ and assume that it can be written as a function of another parameter $s\left(=-\frac{1}{V}\frac{d V}{d \phi}\right)$.
In this case, the dynamical systems become four dimensional and consequently the analysis is slightly more complicated if compared to the case of exponential potential.
In order to investigate the cosmological dynamics for the general scalar field potential, we use the method introduced in \cite{Fang:2008fw}.
We shall see that there are some critical points which exist for a general potential but which do not exist in the exponential potential case.
This type of generalization has been done earlier in the context of braneworld theories \cite{Leyva:2009zz,Escobar:2011cz,Escobar:2012cq,Dutta:2016dnt}, tachyon field \cite{Quiros:2009mz,Fang:2010zze,Farajollahi:2011ym}, quintom field \cite{Leon:2012vt}, $k$-essence \cite{Dutta:2016bbs} and loop quantum gravity \cite{Xiao:2011nh}.
Moreover, for this type of analysis beyond the exponential potential, non-hyperbolic points (critical points whose stability matrix contains a vanishing eigenvalue) are usually obtained.
For this type of points, linear stability theory fails and other complicated mathematical tools, such as Lyapunov function or center manifold theory \cite{Wiggins,Perko,Boehmer:2011tp,Fang:2008fw,TamaniniPhDthesis}, or numerical methods, like perturbation analyses near the critical point \cite{Dutta:2016dnt,Dutta:2016bbs,Roy:2015cna}, need to be employed in order to find the asymptotic behaviour.
Moreover, in order to better understand the cosmological dynamics of this particular models (especially regarding non-hyperbolic critical points), in what follows we also consider two concrete potentials as an example: the hyperbolic potential $V=V_0 \sinh^{-\eta}(\lambda\phi)$ and the inverse powerlaw potential $V=\frac{M^{4+n}}{\phi^n}$.

The organization of the paper is the following. In Sec.~\ref{sec:scalar}, we briefly review the theoretical framework of Scalar-Fluid theories following \cite{Boehmer:2015kta}. In Sec.~\ref{sec:alg}, we present the basic cosmological equations of the model and the formation of an autonomous system of differential equations. In Secs.~\ref{sec:model1} and \ref{sec:model2}, we consider two models corresponding to two distinct algebraic couplings and investigate their dynamics using dynamical systems technique.
For the first interacting model given in Sec.~\ref{sec:model1}, we present two subsections where we focus on two distinct values for one of the parameters.
In both subsections, we consider the two specific potentials mentioned above as examples in order to understand in details the cosmological dynamics in these situations.

{\it Notation}: In this work, we consider the $(-,+,+,+)$ signature convention for the metric tensor. We also consider units where $8\pi G=c=\hslash=1$. The comma notation denotes partial derivatives (i.e.~$\phi_{,\mu}=\partial_{\mu}\phi$).


\section{Scalar Fluid Theories: action and field equations}
\label{sec:scalar}

In this section we briefly present the action and field equations of Scalar-Fluid theories without entering in further theoretical details. The reader interested in more information and applications of Scalar-Fluid theories can refer to \cite{Boehmer:2015kta,Boehmer:2015sha,Koivisto:2015qua,Boehmer:2015ina,Brax:2015fcf,Tamanini:2016klr}.

The total action of Scalar-Fluid theories is given by
\begin{equation}\label{action}
S=\int d^4x\left[\mathcal{L}_{\rm grav}+\mathcal{L}_{\rm mat}+\mathcal{L}_{\phi}+\mathcal{L}_{\rm int}\right],
\end{equation}
where $\mathcal{L}_{\rm grav}$ denotes the gravitational Lagrangian, $\mathcal{L}_{\rm mat}$ denotes the matter Lagrangian, $\mathcal{L}_{\phi}$ denotes the scalar field Lagrangian and $\mathcal{L}_{\rm int}$ denotes the interacting Lagrangian.
The gravitational sector $\mathcal{L}_{\rm grav}$ is given by the usual Einstein-Hilbert Lagrangian
\begin{equation}\label{Lgrav}
\mathcal{L}_{\rm grav}=\frac{1}{2} \sqrt{-g} R,
\end{equation}
where $g$ is the determinant of the metric $g_{\mu\nu}$ and $R$ is the Ricci scalar. The matter Lagrangian $\mathcal{L}_{\rm mat}$ for relativistic fluid described in \cite{Brown:1992kc} is given by
\begin{equation}\label{Lmat}
\mathcal{L}_{\rm mat}=-\sqrt{-g} \rho(\mathfrak{n},\mathfrak{s})+J^{\mu}\left(\varphi_{,\mu}+\mathfrak{s}\, \theta_{,\mu}+\beta_A \alpha^A_{,\mu}\right),
\end{equation}
where $\rho(\mathfrak{n},\mathfrak{s})$ is the energy density of the fluid, assuming that it depends on the particle number density $\mathfrak{n}$ and the entropy density per particle $\mathfrak{s}$. Here $\theta,\,\varphi$ and $\beta_A$ are Lagrange multipliers with $A=1,\,2,\,3$ and $\alpha_A$ are the Lagrangian coordinates of the fluid. The vector density particle number $J^{\mu}$ is related to $\mathfrak{n}$ as
\begin{equation}\label{Jmun}
J^{\mu}=\sqrt{-g}\,\mathfrak{n}\,u^{\mu},\quad |J|=\sqrt{-g_{\mu\nu} J^{\mu}J^{\nu}},\quad \mathfrak{n}=\frac{|J|}{\sqrt{-g}},
\end{equation}
where $u^{\mu}$ is the fluid 4-velocity satisfying $u_{\mu}u^{\mu}=-1$.
The scalar field Lagrangian $\mathcal{L}_{\phi}$ is taken in its canonical form
\begin{equation}\label{Lphi}
\mathcal{L}_{\phi}=-\sqrt{-g}\left[\frac{1}{2}\partial_{\mu}{\phi}\partial^{\mu}{\phi}+V(\phi)\right],
\end{equation}
where $V$ denotes an arbitrary potential for the scalar field $\phi$. Finally, we have to determine the interacting Lagrangian $\mathcal{L}_{\rm int}$. In this work, we consider an algebraic  coupling between the fluid and the scalar field of the type
\begin{align}\label{algLag}
\mathcal{L}_{\rm int}=-\sqrt{-g}f(\mathfrak{n},\mathfrak{s},\phi),
\end{align}
where $f(\mathfrak{n},\mathfrak{s},\phi)$ is an arbitrary function. This type of coupling has been studied in  \cite{Boehmer:2015kta, Tamanini:2016klr} and can lead to late time accelerated scaling solutions similar to the ones obtained in standard interacting models constructed in the past \cite{Amendola:1999er,Zimdahl:2001ar,Mangano:2002gg,Chimento:2003iea}.

Variation of (\ref{action}) with respect to $g_{\mu\nu}$ yields the following Einstein field equations
\begin{equation}\label{EFE}
G_{\mu\nu}=T_{\mu\nu}+T_{\mu\nu}^{(\phi)}+T_{\mu\nu}^{(\rm int)},
\end{equation}
where
\begin{equation}\label{matTmunu}
T_{\mu\nu}=pg_{\mu\nu}+(\rho+p)u_{\mu}u_{\nu},
\end{equation}
\begin{equation}\label{matTphi}
T_{\mu\nu}^{(\phi)}=\partial_{\mu}\phi \partial_{\nu}\phi-g_{\mu\nu}\left[\frac{1}{2}\partial_{\mu}\phi\partial^{\mu}\phi+V(\phi)\right],
\end{equation}
\begin{equation}\label{matTmunuint}
T_{\mu\nu}^{\rm int}=p_{\rm int}g_{\mu\nu}+(\rho_{\rm int}+p_{\rm int})u_{\mu}u_{\nu},
\end{equation}
are the fluid energy momentum tensor, the scalar field energy momentum tensor and the interacting energy momentum tensor, respectively. In the above, the fluid pressure is defined as
\begin{equation}\label{pressure}
p=\mathfrak{n}\frac{\partial \rho}{\partial \mathfrak{n}}-\rho.
\end{equation}
whereas $\rho_{\rm int}$ and $p_{\rm int}$ are the interacting energy density and pressure respectively defined as
 \begin{align}
\rho_{\rm int}=f(\mathfrak{n},\mathfrak{s},\phi),\qquad p_{\rm int}=\mathfrak{n}\frac{\partial f(\mathfrak{n},\mathfrak{s},\phi)}{\partial \mathfrak{n}}-f(\mathfrak{n},\mathfrak{s},\phi).
\end{align}
Varying the action (\ref{action}) with respect to the scalar field yields the modified Klein-Gordon equation
\begin{align}
\Box\phi-\frac{\partial V}{\partial\phi}-\frac{\partial f}{\partial\phi} =0,
\label{Klein_Alg}
\end{align}
where $\Box= \nabla^\mu \nabla_\mu$ and $\nabla_\mu$ is the covariant derivative with respect to the metric $g_{\mu\nu}$.


\section{Basic Cosmological equations}
\label{sec:alg}

In this section, we will consider the cosmological evolution of the Universe based on the interacting model considered in Sec.~\ref{sec:scalar}. As favoured by astronomical observations, we shall consider a spatially flat, homogeneous and isotropic Friedmann-Robertson-Walker (FRW) universe \cite{Ade:2013zuv,Ade:2015xua,Miller:1999qz}, described by the metric
 \begin{align}
 ds^2=-dt^2+a^2(t)(dx^2+dy^2+dz^2),
\end{align}
where $a(t)$ is the scale factor, $t$ is the coordinate time and $x$, $y$, $z$ are Cartesian coordinates.

 Applying this metric to the Einstein field equations (\ref{EFE}) and the Klein-Gordon equation (\ref{Klein_Alg}) yields
\begin{align}\label{Fried_Alg}
3H^2=\left(\rho+\frac{1}{2}\dot{\phi}^2+V+f\right),
\end{align}
\begin{align}
2 \dot{H}+3 H^2=-\left(p+\frac{1}{2}\dot{\phi}^2-V+p_{\rm int}\right),
\end{align}
and
\begin{align}\label{Klein_Alg2}
\ddot{\phi}+3H\dot{\phi}+\frac{\partial V}{\partial \phi}+\frac{\partial f}{\partial \phi}=0,
\end{align}
respectively, where $H=\frac{\dot{a}}{a}$ is the Hubble parameter. It can be shown that at the background level the interaction does not modify the equation of motion of matter \cite{Boehmer:2015kta}
\begin{align}
\dot{\rho}+3H(\rho+p)=0,
\end{align}
where $\rho$ and $p$ denote the energy density and pressure of the fluid with a linear equation of state (EoS) $w$ defined by $p=w\rho$ ($-1\leq w \leq 1$). \\
The effective EoS $w_{\rm eff}$ is defined as
\begin{align}
w_{\rm eff}=\frac{p_{\rm eff}}{\rho_{\rm eff}}=\frac{p+\frac{1}{2}\dot{\phi}^2-V+p_{\rm int}}{\rho+\frac{1}{2}\dot{\phi}^2+V+f} \,.
\end{align}
The Universe undergoes accelerated expansion if the condition $w_{\rm eff}<-\frac{1}{3}$ is satisfied.

As in \cite{Boehmer:2015kta}, in order to convert the cosmological equations (\ref{Fried_Alg})-(\ref{Klein_Alg2}) into an autonomous system of equations, we introduce the following dimensionless variables
\begin{align}
\sigma=\frac{\sqrt{\rho}}{\sqrt{3}H}\,, \quad x=\frac{\dot\phi}{\sqrt{6}H}\,,\quad
y=\frac{\sqrt{V}}{\sqrt{3}H}\,,\quad
z=\frac{ f}{3H^2}\,,\quad s=-\frac{1}{V}\frac{d V}{d \phi}.
\label{variable1}
\end{align}
Here the variable $s$ accounts for the arbitrariness of the self-interacting potentials \cite{Steinhardt:1999nw,delaMacorra:1999ff,Ng:2001hs,Zhou:2007xp,Fang:2008fw,Matos:2009hf,UrenaLopez:2011ur,TamaniniPhDthesis}. Using the dimensionless variables (\ref{variable1}), the Friedmann equation (\ref{Fried_Alg}) becomes
\begin{align}
1=\sigma^2+x^2+y^2+z.
\label{constraint_alg}
\end{align}
This serves as a constraint equation for the phase space, effectively reducing its dimension by one.
Using the above dimensionless variables (\ref{variable1}), the cosmological equations (\ref{Fried_Alg})-(\ref{Klein_Alg2}) can be recast into the following autonomous system of equations
\begin{align}
  x' &= -\frac{1}{2} \left(3x \left((w+1) y^2+w z-w+1\right) + 3 (w-1) x^3-\sqrt{6} s  y^2\right) + x A -B, \label{x_alg}\\
  y' &= -\frac{1}{2} y \left( 3 (w-1) x^2 + 3 \left((w+1) y^2+w z - w-1\right)+\sqrt{6} s  x\right) + y A, \label{y_alg}\\
  z' &= 2 A (z-1)+2 B x-3 z \left((w-1) x^2 + (w+1) y^2 + w (z-1)\right),\label{z_alg}\\
  s'&=-\sqrt{6}\, x\, g(s).\label{s_alg}
\end{align}
where $g(s)=s^2(\Gamma(s)-1)$ and
\begin{align}\label{Gamma}
\Gamma=V\frac{d^2V}{d \phi^2}\left(\frac{d V}{d \phi}\right)^{-2}.
\end{align}
In Eqs.~(\ref{x_alg})-(\ref{s_alg}), we have defined
\begin{align}
A=\frac{p_{\rm int}}{2 H^2} \quad\mbox{and}\quad B=\frac{1}{\sqrt{6} H^2}\frac{\partial\rho_{\rm int}}{\partial\phi} \,.
\end{align}

Different types of potentials lead to different forms of $\Gamma$, which we assume to be a function of $s$. Note that the following analysis is applicable only to potentials where $\Gamma$ can be written as a function of $s$.
If this is not the case then more complicated dynamical systems analysis are needed, usually requiring the addition of further dimensionless variables; see e.g.~\cite{Xiao:2011nh,Barrow:1995xb,Parsons:1995ew,Lazkoz:2007mx,Nunes:2000yc}.
In general if $\Gamma$ is a function of $s$ then scaling solutions naturally appear in the phase space \cite{Nunes:2000yc}, the simplest case being $\Gamma=1$ which corresponds to the case of exponential potential.
In Eqs. (\ref{x_alg})-(\ref{s_alg}) and throughout, a prime denotes differentiation with respect to the number of $e$-folds $N$ defined as $dN=H dt$. It can be seen from Eq.~(\ref{s_alg}) that $s'=0$ only when either $x=0$ or $s=0$ or $\Gamma(s)=1$. We also notice that for $x\neq 0$ it is possible to obtain $s'\neq 0$ and $ g(s)\neq 0$ even when $s=0$, since a particular potential could render the combination $g(s)=s^2(\Gamma(s)-1)$ different from zero. Hence, the necessary condition that $s'=0$ when $x\neq 0$ is $g(s)=0$.

\begin{table}
\begin{tabular}{|c|c|c|c|c|}
\hline
\mbox{} & $\rho_{\rm int}$ & $p_{\rm int}$ & $A$ & $B$ \\
\hline
Model I & $\gamma\,\rho^\alpha\exp(-\beta\phi)$ & $[\alpha(w+1)-1]\rho_{\rm int}$ & $\frac{3}{2}[\alpha(w+1)-1]z$ & $-\beta\sqrt{\frac{3}{2}}z$ \\
\hline
Model II &
$\gamma\phi\rho$ & $w\rho_{\rm int}$ & $\frac{3}{2}wz$ & $\gamma\sqrt{\frac{3}{2}}\left(1-x^2-y^2-z\right)$\\
\hline
\end{tabular}
\caption{Explicit forms of $A$ and $B$ for given $\rho_{\rm int}$. Here $\alpha$, $\beta$ and $\gamma$ are dimensionless parameters.}
\label{models}
\end{table}

In order to close the system, we must specify the function $\rho_{\rm int}$ from which the quantities $A$ and $B$ can be obtained. Specific choices of $\rho_{\rm int}$ let the quantities $A$ and $B$ to depend on $x$, $y$, $z$ only and the resulting system is thus closed without the addition of further dynamical variables.
On the other hand if $A$ and $B$ do not depend solely on $x$, $y$, $z$ then additional extra variables are required, increasing in this manner the dimension of the system.
In what follows, we consider the two choices of $\rho_{\rm int}$ given in Table~\ref{models}, as already studied in \cite{Boehmer:2015kta} for the case of the exponential potential.
For such choices of $\rho_{\rm int}$, it can be seen that $A$ and $B$ depend only on $x$, $y$, $z$; see Table~\ref{models}. From the mathematical point of view, they are simple to analyse and from a physical point of view they are sufficiently complicated to lead to a new and rich cosmological dynamics.
Moreover, as shown in \cite{Brax:2015fcf}, Model~I nicely generalizes the well known chameleon coupling used to screen scalar fields at Solar System scales, while Model~II represents the simplest linear coupling between $\phi$ and $\rho$ one can think of.
As mentioned before different choices from the ones in Table~\ref{models} in general require the introduction of further dimensionless variables: for example simply taking $\rho_{\rm int} = \gamma\phi\rho^2$ would yield $A = \gamma \rho^2 / (\sqrt{6} H^2) = 3\sqrt{3/2} \,\gamma \sigma^4 H^2$, which cannot be rewritten in terms of the variables \eqref{variable1}, but it can be analysed introducing the (compact) variable $u = H_0 / (H + H_0)$; c.f.~\cite{Boehmer:2009tk,Tamanini:2013ltp}. 

Finally we note that from the physical condition $\rho\geq 0$, one has $\sigma^2 \geq 0$, so from Eq.~(\ref{constraint_alg}) one obtains the constraint equation
\begin{align}
x^2+y^2+z\leq 1.
\end{align}
Thus, the four dimensional phase space of the system (\ref{x_alg})-(\ref{s_alg}) is given by
\begin{equation}
\Psi=\left\lbrace(x,y,z)\in \mathbb{R}^3:0\leq x^2+y^2+z\leq 1\right\rbrace \times \left\lbrace s \in \mathbb{R}\right\rbrace.
\end{equation}


\section{Model I}\label{sec:model1}
This section deals with the phase space analysis of the dynamical system~(\ref{x_alg})-(\ref{s_alg}) for the Model I, as given in Table~\ref{models}.
In terms of dimensionless variables (\ref{variable1}), an effective EoS parameter $w_{\rm eff}$ is given by
\begin{align}\label{weff1}
  w_{\rm eff} &= \frac{p_{\rm eff}}{\rho_{\rm eff}} =
  w - (w-1)x^2 - (1+w)y^2 + (1+w) (\alpha-1) z.
\end{align}
We note that the system (\ref{x_alg})-(\ref{s_alg}) is invariant with respect to the transformation $y\rightarrow -y$. So we will restrict the analysis only with positive values of $y$. In general, there are up to nine critical points of the system, depending on the values of parameters $\alpha$, $\beta$, $w$ and $s_*$ as given in Table~\ref{tab:alg_A}. The corresponding eigenvalues of all critical points are given in Table \ref{tab:eigen_alg_A}. In all cases, $s_*$ represents a solution of the equation $g(s)=0$ and $dg(s_*)$ is the derivative of $g(s)$ evaluated at $s=s_*$.

Critical point $O$ does not depend on the potential for its existence and its stability ($s$ is arbitrary). Critical points $A_{1\pm}$, $A_2$, $A_3$, $A_4$, $A_5$ and $A_6$ depend on the particular potential under consideration and there is a copy of each of these point for each solution of $g(s)=0$, i.e.~for each $s_*$. On the other hand, critical point $A_7$ corresponds to the case where the potential is constant, as the $\phi$-derivative of the potential vanishes. It however depends on the concrete form of the potential for its stability. Moreover, it can be seen that point $A_7$ is a special case of point $A_3$ when $s_*=0$.

\begin{table}[b]
\centering
Here: $\Xi=\frac{-2\beta^2+3(\alpha(w+1)-1)(\alpha(w+1)-2)}{3(\alpha(w+1)-2)}$
\begin{adjustbox}{width=1\textwidth}
\small
\begin{tabular}{ccccccc}
  \hline\hline
  Point~ & ~~~~$x$~~~~ & ~~~~$y$ ~~~~& ~~~~$z$~~~~ &~~$s$~~&Existence&~~$w_{\rm eff}$\\
  \hline
  $O$ & 0 & 0 & 0 &$s$&Always&$w$\\

  $A_{1 \pm}$ & $\pm 1$ & 0 & 0&$s_*$ &Always&1\\

  $A_2$ & $\sqrt{\frac{3}{2}} \frac{(1+w)}{s_*}$ &
  $\sqrt{\frac{3}{2}} \frac{\sqrt{(1+w)(1-w)}}{s_*}$ & $0$ &$s_*$&$s_*^2\geq 3(1+w)$&$w$\\

  $A_3$ & $\frac{s_*}{\sqrt{6}}$ & $\sqrt{1-\frac{s_*^2}{6}}$ &
  $0$&$s_*$ &$s_*^2 \leq 6$&$\frac{s_*^2-3}{3}$\\

  $A_4$ & $-\sqrt{\frac{2}{3}} \frac{\beta}{\alpha(w+1)-2}$ &
  0 & $1 - \frac{2\beta^2}{3(\alpha+\alpha w-2)^2}$ &$s_*$&Always&$\Xi$\\[2.5ex]
   $A_5$ & $\sqrt{\frac{3}{2}} \frac{(1+w)(1-\alpha)}{\beta}$ &
  0 & $\frac{3}{2} \frac{(1-\alpha)(1+w)(1-w)}{\beta^2}$&$s_*$&$0 \leq \frac{3(\alpha-1)(w+1)(\alpha(w+1)-2)}{2 \beta^2} \leq 1$&$w$\\[2.5ex]

  $A_6$ & $\sqrt{\frac{3}{2}} \frac{(1+w)\alpha}{s_*-\beta}$ &
  $\frac{\sqrt{6(w+1)\alpha-3(w+1)^2\alpha^2+2\beta(\beta-s_*)}}{\sqrt{2}|s_*-\beta|}$ &
  $\frac{s_*(s_*-\beta)-3(1+w)\alpha}{(\beta-s_*)^2}$&$s_*$&$0 \leq 2\beta(\beta-s_*)$&$-1-\frac{\alpha s_*(w+1)}{\beta-s_*}$\\&&&&&$-3\alpha(w+1)(\alpha(w+1)-2)$\\[2.5ex]

   $A_7$ & $0$ & $1$ &
  $0$&$0$ &Always&$-1$\\

  \hline\hline
\end{tabular}
\end{adjustbox}
\caption{Critical points of Model I.} 
\label{tab:alg_A}
\end{table}

\begin{table}
\centering
Here we have defined: $\Delta^{*}_{\pm}=-\frac{3}{4}(1-w)\left[1\pm\sqrt{\frac{24(1+w)}{s_*^2}-\frac{(7+9w)}{(1-w)}}\right]$\\
 $\Theta_{\pm}=\frac{3}{4}\left[(w-1)\pm  \frac{1}{\beta} \left((w-1) 12\,{\alpha}^{3}{w}^{3}+36\,{\alpha}^{3}{w
}^{2}-24\,{\alpha}^{2}{w}^{3}+36\,{\alpha}^{3}w-96\,{\alpha}^{2}{w}^{2
}-8\,\alpha\,{\beta}^{2}w+12\,\alpha\,{w}^{3}\right.\right.$\\$
\left.\left.+12\,{\alpha}^{3}-120\,{
\alpha}^{2}w-8\,\alpha\,{\beta}^{2}+84\,\alpha\,{w}^{2}+9\,w{\beta}^{2
}-48\,{\alpha}^{2}+132\,\alpha\,w+7\,{\beta}^{2}-24\,{w}^{2}+60\,
\alpha-48\,w-24 \right)^{1/2}
 \right]$\\[1.5ex]

$\Lambda^{*}_{\pm}=\frac{1}{4(\beta-s_*)}\left[-3\,\alpha\,s_{*}\,(w+1)-6\,(\beta-s_{*})\pm
\left\lbrace-216\,{\alpha}^{3}{w}^{3}-72\,{\alpha}^{2}\beta\,s_{*}\,{w}^{2}
+81\,{\alpha}^{2}{s_{*}}^{2}{w}^{2}-648\,{\alpha}^{3}{w}^{2}-144\,{
\alpha}^{2}\beta\,s_{*}\,w\right.\right.$\\$\left.\left.+162\,{\alpha}^{2}{s_{*}}^{2}w-648\,{\alpha}
^{3}w-72\,{\alpha}^{2}\beta\,s_{*}+81\,{\alpha}^{2}{s_{*}}^{2}+432\,{
\alpha}^{2}{w}^{2}+144\,\alpha\,{\beta}^{2}w+36\,\alpha\,\beta\,s_{*}
\,w-180\,\alpha\,{s_{*}}^{2}w+48\,{\beta}^{3}s_{*}\right.\right.$\\$\left.\left.-96\,{\beta}^{2}{s_{
*}}^{2}+48\,\beta\,{s_{*}}^{3}-216\,{\alpha}^{3}+864\,{\alpha}^{2}w+
144\,\alpha\,{\beta}^{2}+36\,\alpha\,\beta\,s_{*}-180\,\alpha\,{s_{*}}
^{2}+432\,{\alpha}^{2}+36\,{\beta}^{2}-72\,\beta\,s_{*}+36\,{s_{*}}^{2
}\right\rbrace^{1/2}\right]$
\begin{adjustbox}{width=1\textwidth}
\small
\begin{tabular}{ccccccc}
\hline\hline
Point &$E_1$ & $E_2$ &$E_3$ &$E_4$ & Stability \\
\hline
$O$ & $0$ & $\frac{3}{2}(w-1)$ &$\frac{3}{2}(w+1)$ &$3(w+1)(1-\alpha)$& Saddle\\[1.5ex]
$A_{1 \pm}$&$3(1-w)$&$3 \mp \frac{\sqrt{6}}{2}s_*$&$\sqrt{6}(\sqrt{6}\mp \beta)-3\alpha(w+1)$&$\mp \sqrt{6}\,dg(s_*)$&Unstable node/Saddle\\ [1.5ex]

$A_2$ & $-\frac{3(\beta+s_*(\alpha-1))(w+1)}{s_*}$ & $\Delta^{*}_+$ &$\Delta^{*}_-$&$-\frac{3(w+1)\,dg(s_*)}{s_*}$&  Figs. \ref{fig:C_G_stable_alpha1}, \ref{fig:E_G_stable_alpha3}.\\[1.5ex]
$A_3$ & $\frac{s_*^2}{2}-3$ & $s_*^2-3(w+1)$ & $s_*(s_*-\beta)-3\alpha(w+1)$ & $-s_* dg(s_*)$&Stable node/Saddle\\[4ex]
&&&&& Unstable node/Saddle\\
$A_4$&$\frac{-2\beta^2+3(\alpha(w+1)-2)^2}{\alpha(w+1)-2}$&$\frac{-2\beta^2+3(w+1)(\alpha-1)(\alpha(w+1)-2)}{\alpha(w+1)-2}$&$\frac{1}{2}\frac{2\beta(s_*-\beta)+3\alpha(w+1)(\alpha(w+1)-2)}{\alpha(w+1)-2}$&$\frac{2\beta\, dg(s_*)}{\alpha(w+1)-2}$& ($\alpha=1$)\\

&&&&& See Fig. \ref{fig:E_G_stable_alpha3} for $\alpha=3$\\[4ex]
$A_5$&$\frac{3}{2}\frac { \left( w+1 \right)  \left( \alpha\,s_{*}+\beta-s_{*}
 \right) }{\beta}
$&$\Theta_{+}$&$\Theta_{-}$&$\frac{3(\alpha-1)(w+1) dg(s_*)}{\beta}$&Saddle (for $w=0$)\\[2.5ex]

$A_6$&$\frac{-3(\beta+(\alpha-1) s_*)(w+1)}{(\beta-s_*)}$&$\Lambda^{*}_{+}$&$\Lambda^{*}_{-}$&$\frac{3\alpha (w+1) dg(s_*)}{\beta-s_*}$& Figs. \ref{fig:C_G_stable_alpha1}, \ref{fig:E_G_stable_alpha3}.\\[4.5ex]

 &&&&&Stable if $g(0)>0$, $\alpha>0$\\
 $A_7$  & $-3(w+1)$ & $-3\alpha(w+1)$ & $\frac{1}{2}\left(-\sqrt{12 g(0)+9}-3\right)$& $\frac{1}{2}\left(\sqrt{12 g(0)+9}-3\right)$&Saddle if $g(0)<0$\\
&&&&& or $\alpha<0.$\\
 &&&&&See App.~if $g(0)=0$\\
\hline\hline

\end{tabular}
\end{adjustbox}
\caption{Stability of critical points of Model I. } \label{tab:eigen_alg_A}
\end{table}

Due to the complexity of model I, we will consider only two distinct values of $\alpha$: namely $\alpha=1$ or 3. For these two choices of $\alpha$, we can see how these interacting models can well describe the observed late time dynamics of the Universe.

\subsection{The case $\alpha=1$} \label{alpha1}

We will now investigate the phase space analysis for the choice $\alpha=1$. For this particular choice of $\alpha$, critical point $A_5$  coincides with the origin $O$. So the total number of critical points reduces to eight. Critical points $O$, $A_{1\pm}$, $A_4$ exist for any values of parameters $\beta$, $w$, $s_*$. Critical point $A_2$ exists for $s_*^2\geq 3(w+1)$, critical point $A_3$ exists for $s_*^2\leq 6$ and point $A_6$ exists whenever $3(1-w^2)\geq 2\beta(s_*-\beta)$.

Critical point $O$ corresponds to an unaccelerated matter dominated universe ($w_{\rm eff}=w$). It is non-hyperbolic in nature due to the vanishing of at least one of its eigenvalues, however it behaves as a saddle as two of its corresponding non-vanishing eigenvalues are opposite in sign. Points $A_{1 \pm}$ correspond to an unaccelerated, scalar field kinetic energy dominated solution with stiff fluid effective EoS ($w_{\rm eff}=1$). Point $A_{1+}$ is an unstable node whenever $\beta<\frac{\sqrt{6}}{2}(1-w)$, $s_*<\sqrt{6}$ and $dg(s_*)<0$, otherwise it is a saddle, whereas point $A_{1-}$ is an unstable node whenever $\beta>-\frac{\sqrt{6}}{2}(1-w)$, $s_*>-\sqrt{6}$ and $dg(s_*)>0$, otherwise it is a saddle.
Point $A_2$ corresponds to a scaling solution with $w_{\rm eff}=w$ where the energy densities of DM and DE are both non zero. This means that the Universe behaves as if it is dominated by matter completely, even though both DE and DM contribute to the total energy. It is a stable node when $0<\frac{24 (1+w)}{s_*^2}-\frac{(7+9w)}{(1-w)}<1$, $\frac{\beta}{s_*}>0$ and $s_*\,dg(s_*)>0$, it is a stable spiral when $\frac{24 (1+w)}{s_*^2}-\frac{(7+9w)}{(1-w)}<0$, $\frac{\beta}{s_*}>0$ and $s_*\,dg(s_*)>0$, otherwise it is a saddle. Point $A_3$ corresponds to a scalar field dominated universe.
It represents an accelerated universe whenever $s_*^2<2$. It is a stable node when $s_*^2<3(w+1)$, $s_*\beta<s_*^2-3(w+1)$ and $s_*dg(s_*)>0$, otherwise it is a saddle. Point $A_4$ corresponds to a solution where the matter energy density vanishes but the kinetic part of the scalar field and the interacting energy density are non-zero. It is not stable as its eigenvalue $E_2>0$ (it is an unstable node when $2\beta^2>3(1-w)$, $2\beta\,s_*<3(1-w)+2\beta^2$ and $\beta\,dg(s_*)<0$, otherwise it is a saddle). It corresponds to an accelerated universe when $2\beta^2<(3w-1)(1-w)$. Due to the complexity of critical point $A_6$, we will analyse its stability only on a physically interesting case, i.e.~for $w=0$. It corresponds to an accelerated universe when $\frac{s_*}{\beta}>-2$ and it stands for a solution where the matter energy density vanishes but the scalar field and interacting energy density are both non-zero (see \cite{Tamanini:2015iia} for a discussion regarding this kind of solutions). It is either stable or a saddle node depending on the values of $s_*$ and $\beta$.

\begin{figure}
\centering
\includegraphics[width=8cm,height=6cm]{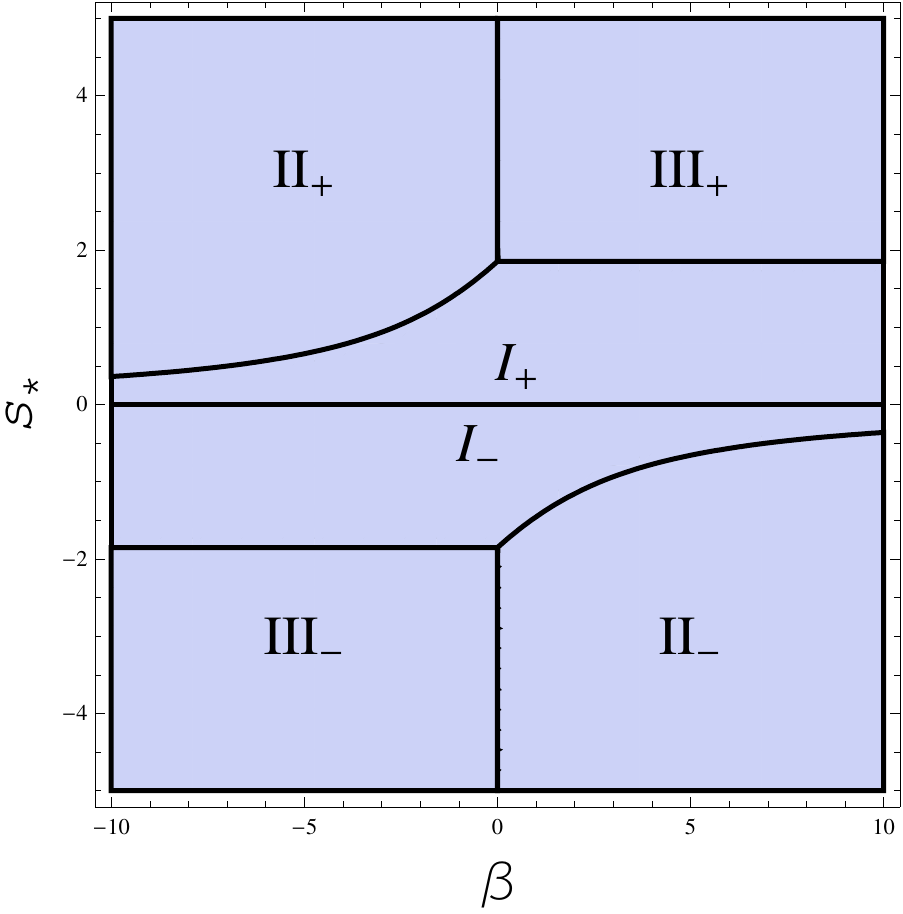}
\caption{Regions of stability of critical points $A_2$, $A_3$ and $A_6$ in the $(\beta,s_*)$ parameter space. Region I$_+$  corresponds to the region of stability of point $A_3$ when $dg(s_*)>0$, whereas region I$_-$  corresponds to the region of stability of point $A_3$ when $dg(s_*)<0$. Region II$_+$ and II$_{-}$ corresponds to the region of stability of point $A_6$ when $dg(s_*)>0$ and $dg(s_*)<0$ respectively. Region III$_{+}$ corresponds to the regions of stability of point $A_2$ when $dg(s_*)>0$ whereas region III$_{-}$ for $dg(s_*)<0$. Here we have taken $w=0$ and $\alpha=1$.}
\label{fig:C_G_stable_alpha1}
\end{figure}

The stability analysis for points $A_2$, $A_3$ and $A_6$ is confirmed numerically as shown in Fig.~\ref{fig:C_G_stable_alpha1} by considering $s_*$ as a parameter without specifying the concrete form of the scalar field potential and by focusing on the most physically interesting case ($w=0$). Critical point $A_7$ corresponds to an accelerated scalar field dominated solution ($w_{\rm eff}=-1$). It is stable whenever $g(0)>0$ and a saddle whenever $g(0)<0$. However, it is a non-hyperbolic point if $g(0)=0$, in which case linear stability theory fails to determine its stability and other more complicated mathematical tools, like center manifold theory, are required \cite{Boehmer:2011tp,Fang:2008fw,TamaniniPhDthesis}. The full analysis of the stability of this point using center manifold theory is given in the appendix A. From that analysis, point $A_7$ corresponds to a late time scalar field dominated attractor if $\Gamma(0)>1$ (c.f.~Eq.~\eqref{Gamma}).

From the above stability analysis, we can see that whenever $\beta$ and $s_*$ are of the same sign (see Fig.~\ref{fig:C_G_stable_alpha1}), the late time behaviour of the universe is undistinguishable from the case of a canonical scalar field without interaction \cite{Fang:2008fw}. In this scenario, the late time universe will correspond to a scalar field dominated point $A_3$ (or $A_7$) or a scaling solution $A_2$. However, whenever $\beta$ and $s_*$ have opposite sign (see Fig.~\ref{fig:C_G_stable_alpha1}), then a new late time behaviour arises. In this case the late universe will either correspond to a scalar field dominated solution $A_3$ (or $A_7$) or to an accelerated solution $A_6$ where the scalar field energy density and the interacting energy density do not vanish, but the matter energy density does vanish.
This solution can thus alleviate the coincidence problem as it corresponds to an accelerated late time attractor where the DE density does not dominate completely.
This is similar to what one can achieve by introducing interacting terms at the level of the field equations \cite{Amendola:1999er,Zimdahl:2001ar,Mangano:2002gg,Chimento:2003iea}.
Note also that points $A_2$, $A_3$ and $A_6$ cannot be late time attractors simultaneously (see Fig.~\ref{fig:C_G_stable_alpha1}). Hence, depending on the choice of parameters, the Universe either evolves from a matter dominated solution to a late time, accelerated, scalar field dominated solution, describing in this way the DM to DE transition, or it reaches an accelerated scaling solution capable of solving the cosmic coincidence problem.

In order to better understand the dynamics of these cosmological models, it is now interesting to focus the analysis above on concrete forms of the scalar field potential, especially given the fact that the stability of the critical points depends strongly on the values of $s_*$ and $dg(s_*)$.
In the following examples we choose two specific potentials and analyze their dynamics in detail.

\subsubsection*{Example I:~$V=V_0\sinh^{-\eta}(\lambda\phi)$}

In this first case, we consider the hyperbolic potential
\begin{align}
V=V_0\sinh^{-\eta}(\lambda\phi)
\end{align}
where $V_0$ and $\lambda$ are parameters with suitable dimensions and $\eta$ is a dimensionless parameter. This potential was first introduced in \cite{Sahni:1999gb}. For a canonical scalar field, it has  been studied using dynamical systems techniques in \cite{Fang:2008fw,Roy:2013wqa}, the cosmological dynamics of some alternative cosmological models using this potential has been also studied in \cite{Leyva:2009zz,Dutta:2016bbs}.
For this potential, we have
\begin{align}
g(s)=\frac{s^2}{\eta}-\eta\lambda^2
\end{align}
so that
\begin{align}
s_*=\pm \eta\lambda,~~dg(s_*)=\frac{2s_*}{\eta}
\end{align}
Critical points $A_{1\pm}$, $A_2$, $A_3$, $A_4$, $A_6$ will have each exactly two copies for two solutions $s_*=\pm \eta\lambda$, point $O$ exists for any arbitrary potential.

\begin{figure}
\centering
\includegraphics[width=8cm,height=6cm]{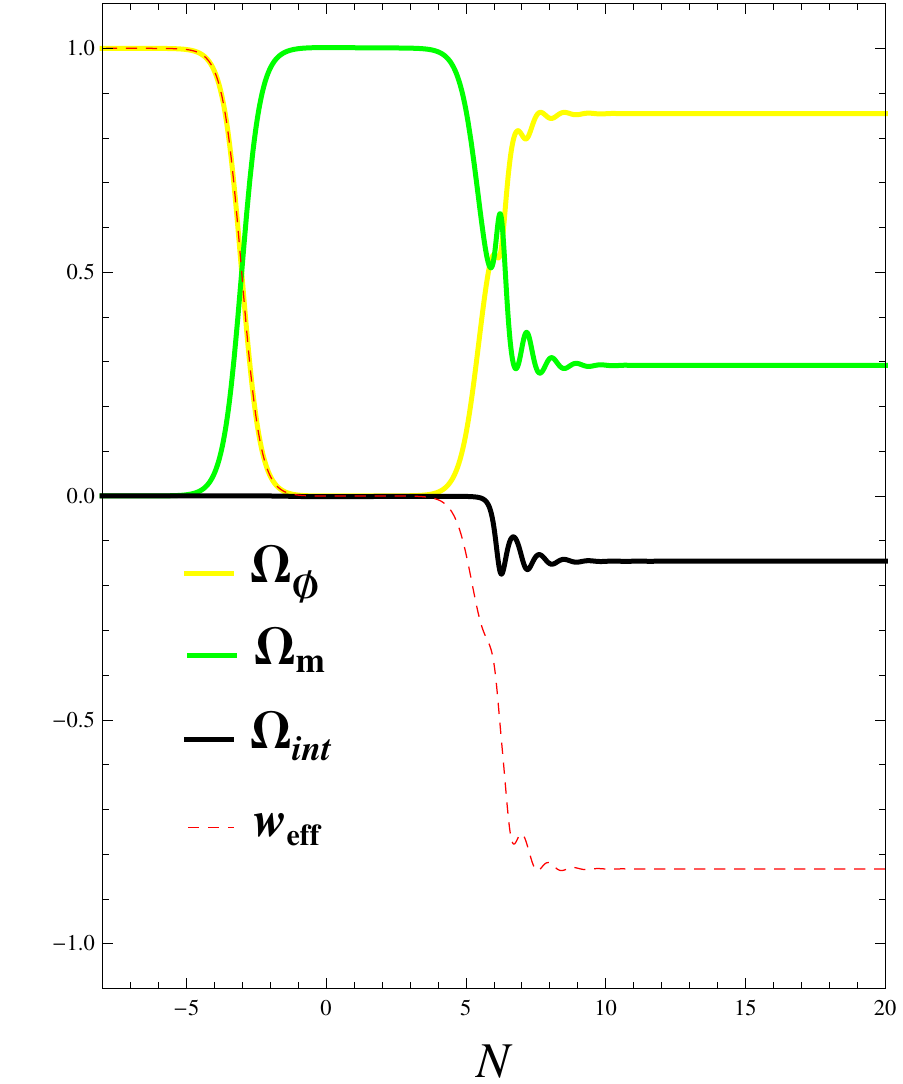}

\caption{Evolution of the relative energy density of DE ($\Omega_\phi$), DM ($\Omega_m$), interaction ($\Omega_{\rm int}$) together with effective equation of state ($w_{\rm eff}$) for interacting model I with potential $V=V_0 \sinh^{-\eta}(\lambda\phi)$. Here we have taken $w=0$, $\alpha=1$, $\beta=10$, $\lambda=-2$, $\eta=1$.
}
\label{fig:weff_omega_m_phi_int_alpha_1}
\end{figure}

As already mentioned above, the properties of the critical point $O$ do not depend on the scalar field potential.
Point $A_{1+}$ is an unstable node whenever $\beta<\frac{\sqrt{6}}{2}(1-w)$, $\eta\lambda<\sqrt{6}$ and $\lambda<0$, otherwise it is a saddle, whereas point $A_{1-}$ is an unstable node whenever $\beta>-\frac{\sqrt{6}}{2}(1-w)$, $\eta\lambda>-\sqrt{6}$ and $\lambda>0$, otherwise it is a saddle.
Point $A_2$ is a stable node when $0<\frac{24 (1+w)}{\eta^2\lambda^2}-\frac{(7+9w)}{(1-w)}<1$, $\frac{\beta}{\eta\lambda}>0$ and $\eta>0$, it is a stable spiral when $\frac{24 (1+w)}{\eta^2\lambda^2}-\frac{(7+9w)}{(1-w)}<0$, $\frac{\beta}{\eta\lambda}>0$ and $\eta>0$, otherwise it is a saddle.
Point $A_3$ corresponds to a scalar field dominated universe. It represents an accelerated universe whenever $\eta^2\lambda^2<2$.
It is a stable node when $\eta^2\lambda^2<3(w+1)$, $\eta\lambda\beta<\eta^2\lambda^2-3(w+1)$ and $\eta>0$, otherwise it is a saddle. Point $A_4$ is an unstable node when $2\beta^2>3(1-w)$, $2\beta\,\eta\lambda<3(1-w)+2\beta^2$ and $\beta\,\lambda<0$, otherwise it is a saddle.
As in the general potential case, for critical point $A_6$ we will consider only the physically interesting value $w=0$.
For this choice $A_6$ represents an accelerated universe when $\frac{\eta\lambda}{\beta}>-2$.
It is either a stable or a saddle node depending on the values of $\eta$, $\lambda$ and $\beta$.
This solution can solve the coincidence problem as it corresponds to an accelerated late time attractor where the DE density does not dominate completely.
For example if we numerically choose 
$\alpha=1,\,\beta=10,\,\lambda=-2,\,\eta=1$, we obtain  $E_1=-2.5$, $E_2=-1.53-8.16\,i $, $E_3=-1.53+8.16\,i$, $E_4=-0.67$, $w_{\rm eff}=-0.83$ and the DE density parameter denoted by $\Omega_{\phi}=x^2+y^2=0.85$. Finally, critical point $A_7$ corresponds to a late time accelerated solution whenever $\eta<0$, and it is a saddle otherwise.

Hence, depending on the choices of the model parameters, the universe starts from a stiff matter dominated solution $A_{1\pm}$ ($w_{\rm eff}=1$) and evolves towards either an accelerated scalar field dominated solution $A_3$ (or $A_7$), an unaccelerated scaling solution $A_2$, or an accelerated scaling solution $A_6$ through a matter dominated solution $O$.
Thus, this model can well describe the deceleration to acceleration transition corresponding to the domination of DE over DM at late times.
Moreover it can solve the cosmic coincidence problem whenever point $A_6$ is the late time attractor, regardless of fine tuning issues concerning the amplitude of the scalar field potential $V_0$ which is allowed to take any positive non-vanishing value.
This behaviour is shown by Fig.~\ref{fig:weff_omega_m_phi_int_alpha_1} where the time evolution of all energy densities involved, together with the effective EoS of the universe, have been plotted choosing initial conditions leading to a long lasting matter dominated phase, as required by cosmological observations.
From the figure it is clear that the final stage of the universe is an accelerated expansion where the energy densities of matter and dark energy remain fixed to the same order of magnitude.
Furthermore during the transition from dark matter to dark energy domination the effective EoS of the universe presents some non-linear behaviour, the ``oscillations'' in Fig.~\ref{fig:weff_omega_m_phi_int_alpha_1}, which distinguish this model from the standard $\Lambda$CDM evolution and could in principle lead to observational signatures to look for in future astronomical observations.


\subsubsection*{Example II:  $V=\frac{M^{4+n}}{\phi^n}$}

In this second example, we consider the potential
\begin{equation}
    V=\frac{M^{4+n}}{\phi^n} \,,
 \end{equation}
where $M$ is a mass scale and $n$ is a dimensionless parameter. This potential can lead to tracking behaviour \cite{Zlatev:1998tr} and the cosmological evolution of some cosmological models where this potential appears have recently been studied using dynamical system technique \cite{Dutta:2016dnt,Dutta:2016bbs}.

\begin{figure}
\centering
\subfigure[]{%
\includegraphics[width=8cm,height=6cm]{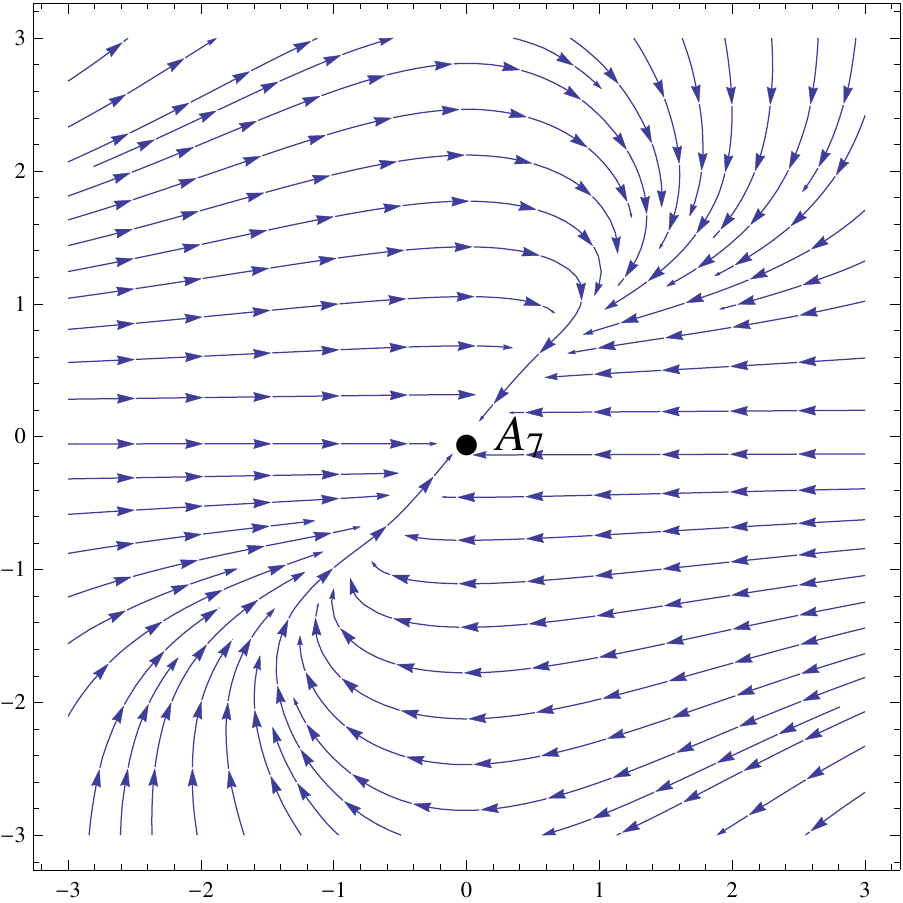}}
\qquad
\subfigure[]{%
\includegraphics[width=8cm,height=6cm]{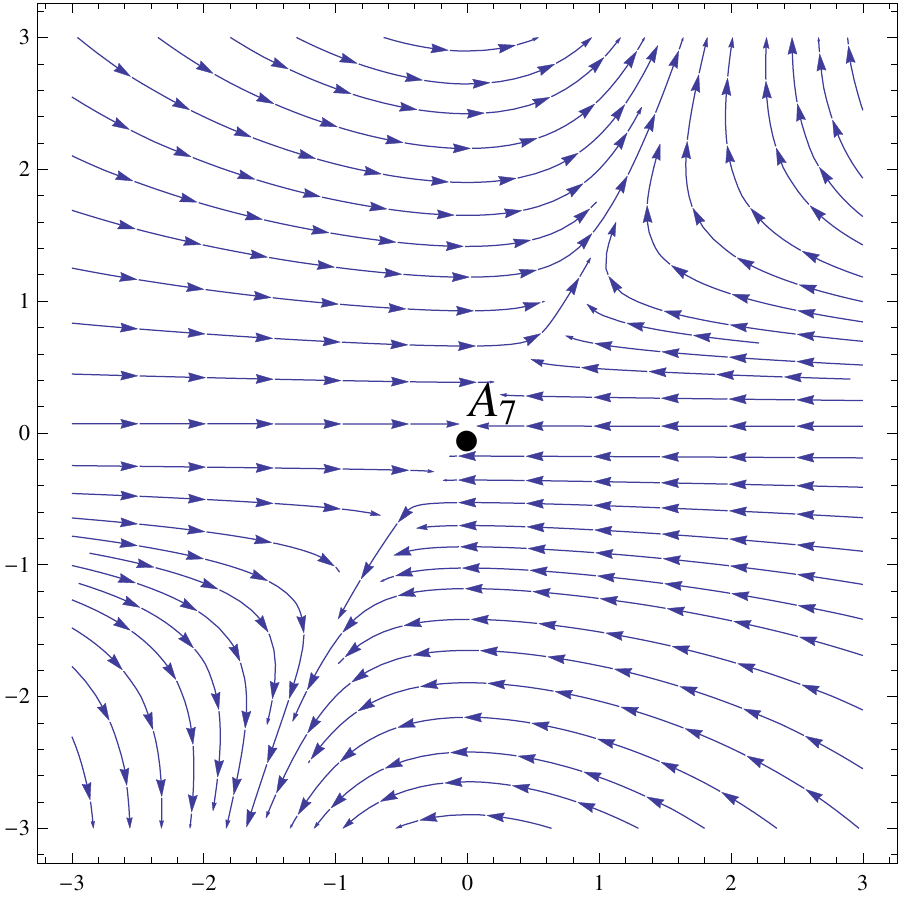}}
\caption{Projection of the system (\ref{x_alg})-(\ref{s_alg}) for model I with potential $V=\frac{M^{4+n}}{\phi^n}$ on the $x-s$ plane. In the panel (a) we take $n=10$ where point $A_7$ is stable, in panel (b) we take $n=-10$ where point $A_7$ is not stable. Here we have taken $w=0$, $\alpha=1$, $\beta=-2$.}\label{fig:stream plot_A7_powerlaw_1}
\end{figure}

In this case, we have simply
\begin{equation}\label{eq30}
g(s)=\frac{s^2}{n} \,,
\end{equation}
implying that $s_*=0$.
This means that critical point $A_2$ does not exist for this potential, and that point $A_3$ reduces to point $A_7$.
Looking at the eigenvalues given in Table \ref{tab:eigen_alg_A} for the case $s_*=0$ and $dg(s_*)=0$, one can immediately realise that all the allowed critical points are non-hyperbolic.
Point $A_{1+}$ behaves as an unstable node whenever $\beta<\frac{\sqrt{6}}{2}(1-w)$, otherwise it behaves as a saddle, whereas point $A_{1-}$ behaves as an unstable node whenever $\beta>-\frac{\sqrt{6}}{2}(1-w)$, otherwise it behaves as a saddle.
Point $A_4$ is not stable as its eigenvalue $E_2>0$ (it behaves as unstable node when $2\beta^2>3(1-w)$, otherwise it behaves as a saddle).
Point $A_6$ behaves as a saddle since the eigenvalue $E_2$ is always positive and $E_1$ is negative.
For critical point $A_7$, eigenvalues $E_1$, $E_2$ are negative whereas $E_3$ and $E_4$ vanish.
Therefore, linear stability theory fails to determine its stability and other complicated mathematical tools like center manifold theory are required.
The full analysis of the stability of this point using center manifold theory is given in the appendix A. From the analysis, point $A_7$ corresponds to a late time scalar field dominated attractor if $\Gamma(0)>1$ (or $n>0$).
From the above analysis we see that for $n>0$ the Universe evolve towards a unique late time attractor $A_7$ through a matter dominated phase $O$.
Thus, we see that for this potential with $n>0$, this cosmological interacting model has only one late time accelerated, scalar field dominated attractor for a wide range of initial conditions.
If instead $n<0$ then there is no finite late time attractor and all trajectories evolve towards some critical point at infinity.
This can be shown for example in the behaviour displayed in Fig.~\ref{fig:stream plot_A7_powerlaw_1}, where point $A_7$ is a global attractor for $n>0$ while it becomes a saddle if $n<0$.


\subsection{The case $\alpha=3$}

\begin{figure}
\centering
\includegraphics[width=8cm,height=6cm]{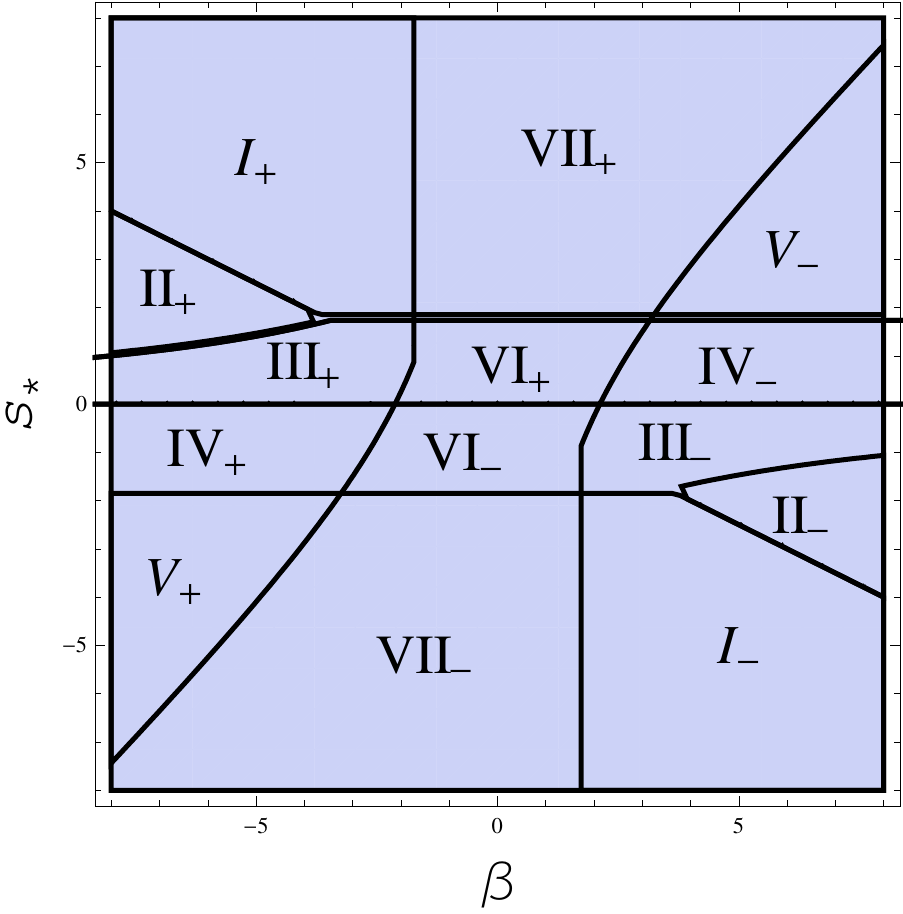}
\caption{Stability region of critical points $A_2$, $A_3$, $A_4$ and $A_6$ in the $(\beta,s_*)$ parameter space. Regions I$_{+}$, V$_{-}$, VII$_{+}$ correspond to the regions of stability of point $A_2$ when $dg(s_*)>0$, regions I$_{-}$, V$_{+}$, VII$_{-}$ correspond to the regions of stability of point $A_2$ when $dg(s_*)<0$. Regions III$_+$, IV$_{-}$ and VI$_{+}$ correspond to the regions where $A_3$ is stable when $dg(s_*)>0$ and regions III$_-$, IV$_{+}$ and VI$_{-}$ correspond to the regions where $A_3$ is stable when $dg(s_*)<0$. Regions I$_{+}$, II$_{+}$, III$_{+}$, IV$_{+}$ and V$_{+}$ correspond to regions where point $A_4$ is stable  when $dg(s_*)>0$ and regions I$_{-}$, II$_{-}$, III$_{-}$, IV$_{-}$ and V$_{-}$  correspond to regions where point $A_4$ is stable when $dg(s_*)<0$. Region II$_{+}$ corresponds to region where point $A_6$ is stable when $dg(s_*)>0$ whereas region II$_{-}$ corresponds to region where point $A_6$ is stable when $dg(s_*)<0$. Here we have assumed $w=0$ and $\alpha=3$.}
\label{fig:E_G_stable_alpha3}
\end{figure}

We will now investigate the case where $\alpha$ is different from $1$. Following \cite{Boehmer:2015kta} we will consider $\alpha=3$ for simplicity of the analysis.
The qualitative properties of critical points $O$, $A_{1 \pm}$, $A_2$, $A_3$, $A_7$ remain roughly the same as the case $\alpha=1$, except a slightly difference in the stability regions of $(\beta,s_*)$ parameter space.
The stability regions of the critical points in the $(\beta,s_*)$ parameter space are shown in Fig.~\ref{fig:E_G_stable_alpha3}.
Point $O$ corresponds again to an unaccelerated matter dominated universe ($w_{\rm eff}=w$). It is still a saddle as two of its corresponding non-vanishing eigenvalues are opposite in sign.
Points $A_{1 \pm}$ correspond to an unaccelerated scalar field kinetic energy dominated solution, with stiff fluid effective EoS ($w_{\rm eff}=1$).
Point $A_{1+}$ is an unstable node whenever $\beta<-\frac{\sqrt{6}}{2}(3w+1)$, $s_*<\sqrt{6}$ and $dg(s_*)<0$, otherwise it is a saddle, whereas point $A_{1-}$ is an unstable node whenever $\beta>\frac{\sqrt{6}}{2}(3w+1)$, $s_*>-\sqrt{6}$ and $dg(s_*)>0$, otherwise it is a saddle.
Point $A_2$ corresponds to a scaling solution with $w_{\rm eff}=w$: the Universe behaves as if it is matter dominated even though both DE and DM contributions are both non-zero.
Point $A_3$ characterises a scalar field dominated universe. It represents an accelerated universe whenever $s_*^2<2$. It is a stable node when $s_*^2<3(w+1)$, $s_*^2<s_*\beta+(9w+1)$ and $s_*dg(s_*)>0$, otherwise it is a saddle.
Point $A_4$ corresponds to a solution where the scalar field kinetic energy and the interacting energy density do not vanish, while the matter energy density instead vanishes.
It stands for an accelerated universe whenever $\frac{2(\beta^2-3)-27w(w+1)}{3w+1}>1$ and it can be stable as illustrated numerically in Fig.~\ref{fig:E_G_stable_alpha3} (at least for $w=0$) as contrary to the case $\alpha=1$.
Point $A_5$ exists only when $\beta^2\geq 3(w+1)(3w+1)$. It corresponds to the solution where the potential energy density vanishes but the scalar field kinetic energy, matter energy density and interacting energy density do not vanish.
It also represents an unaccelerated scaling solution where the Universe expands as if it was matter dominated ($w_{\rm eff}=w$).
This point is saddle (at least for $w=0$). Again, due to the complexity of critical point $A_6$, we will focus only on pressure-less fluid ($w=0$) to analyse its stability.
It corresponds to an accelerated universe when $\frac{s_*}{\beta}>-\frac{2}{7}$, and it defines a solution where the matter energy density vanishes but the scalar field and interacting energy density are both non-zero.
It is either a stable or a saddle node depending on the values of $s_*$ and $\beta$.
This behaviour has been confirmed numerically as shown in Fig.~\ref{fig:E_G_stable_alpha3} by considering $s_*$ as a parameter without specifying the concrete form of the scalar field potentials.
Finally, critical point $A_7$ corresponds to an accelerated scalar field dominated solution ($w_{\rm eff}=-1$). Similarly to the case $\alpha=1$, it is stable whenever $g(0)>0$ and saddle whenever $g(0)<0$. It is a non-hyperbolic point if $g(0)=0$.
Exactly as in the case $\alpha=1 $ this point is stable whenever $\Gamma(0)>1$, as can be verified using center manifold theory.

From the above analysis, we see that the cosmological dynamics is complicated compared to the case of $\alpha=1$. In this case depending on some values of the parameters, we obtain multiple attractors, namely $A_4$ and $A_2$ or $A_4$ and $A_3$ (see Fig.~\ref{fig:E_G_stable_alpha3}). However as in the case $\alpha=1$, this model can successfully describe the late time behaviour of the Universe.  We shall now study the stability of critical points for the same two concrete potentials considered before.

\subsubsection*{Example I: $ V=V_0\sinh^{-\eta}(\lambda\phi)$}\label{subsec:model_I_3}

\begin{figure}
\centering
\includegraphics[width=8cm,height=6cm]{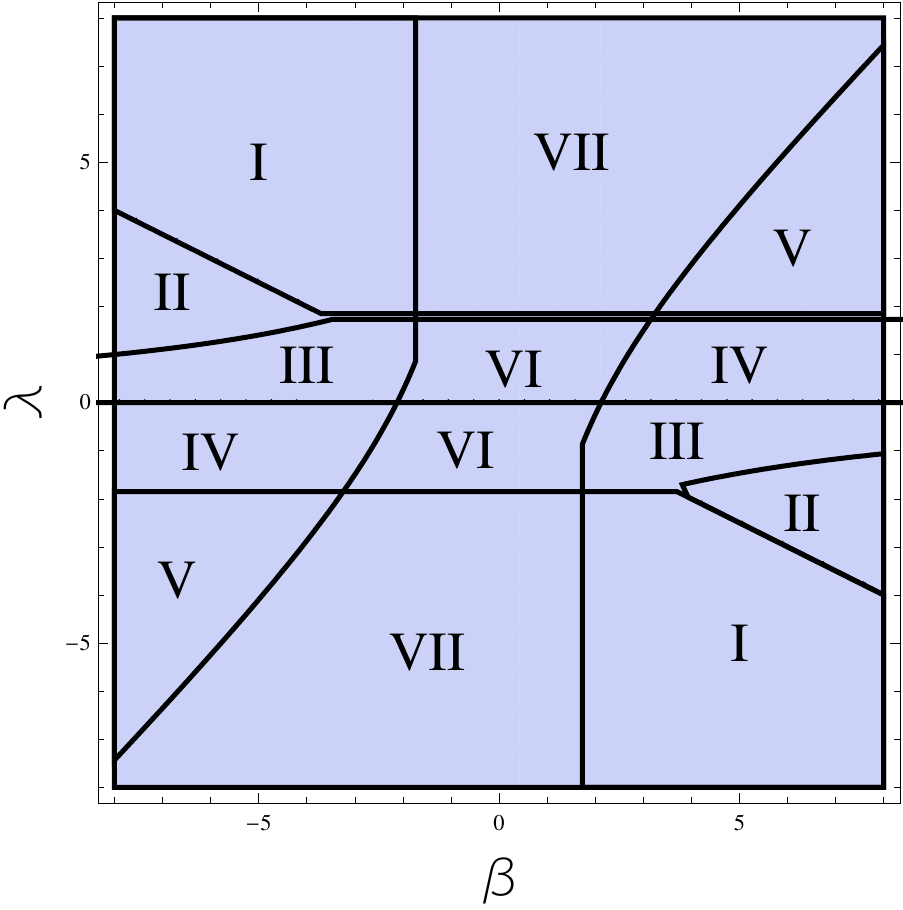}
\caption{Stability region of critical points $A_2$, $A_3$, $A_4$ and $A_6$ in the $(\beta,\lambda)$ parameter space. Regions I, V, VII correspond to the regions of stability of point $A_2$. Regions III, IV and VI correspond to the regions where $A_3$ is stable. Regions I, II, III, IV and V correspond to regions where point $A_4$ is stable. Region II corresponds to region where point $A_6$ is stable. Here we have assumed the potential $ V=V_0\sinh^{-\eta}(\lambda\phi)$ with $w=0$, $\alpha=3$, $\eta=1$.}
\label{region_sinh3}
\end{figure}

\begin{figure}
\centering
\includegraphics[width=8cm,height=6cm]{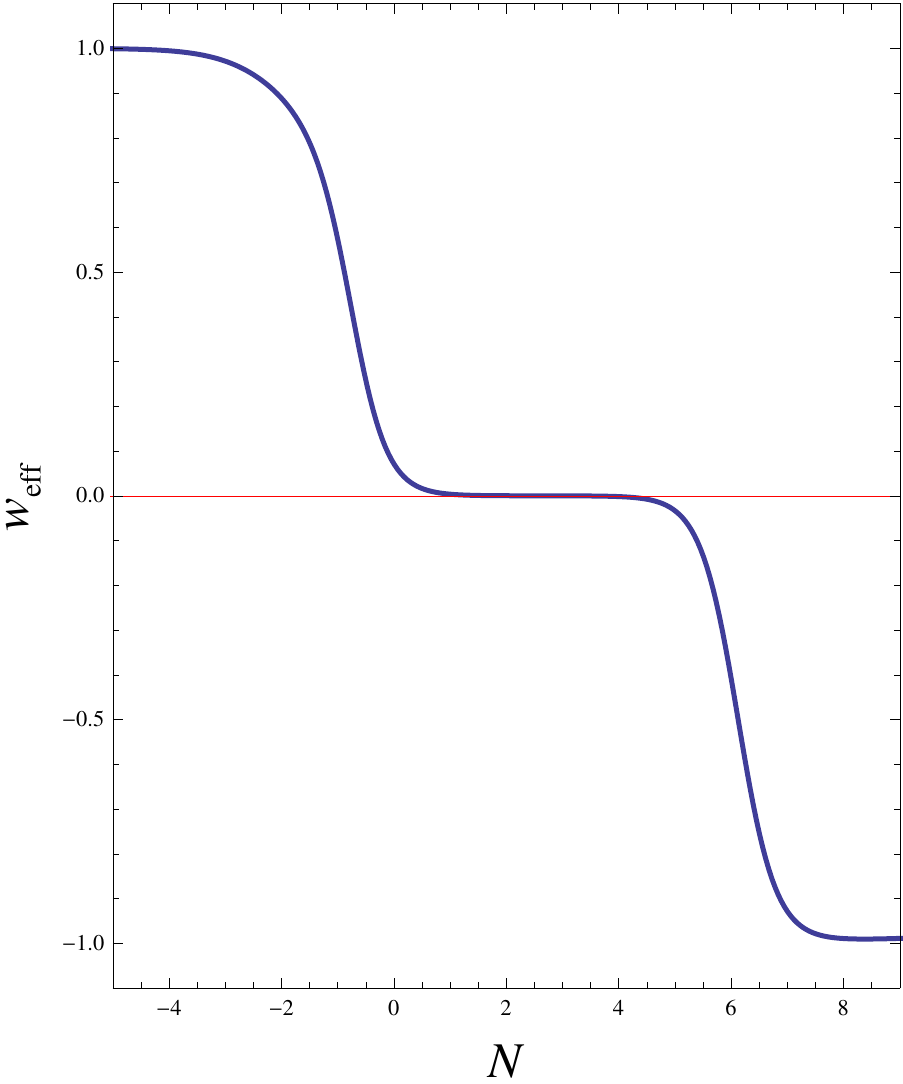}
\caption{Plot of $w_{\rm eff}$ versus $N$ of model I with potential $ V=V_0\sinh^{-\eta}(\lambda\phi)$ showing a transition from matter dominated epoch (point $O$) to a DE dominated epoch (point $A_3$). Here we have taken $w=0$, $\alpha=3$, $\beta=2$, $\lambda=0.5$, $\eta=1$.}
\label{fig:weff_I_3}
\end{figure}

\begin{figure}
\centering
\includegraphics[width=8cm,height=6cm]{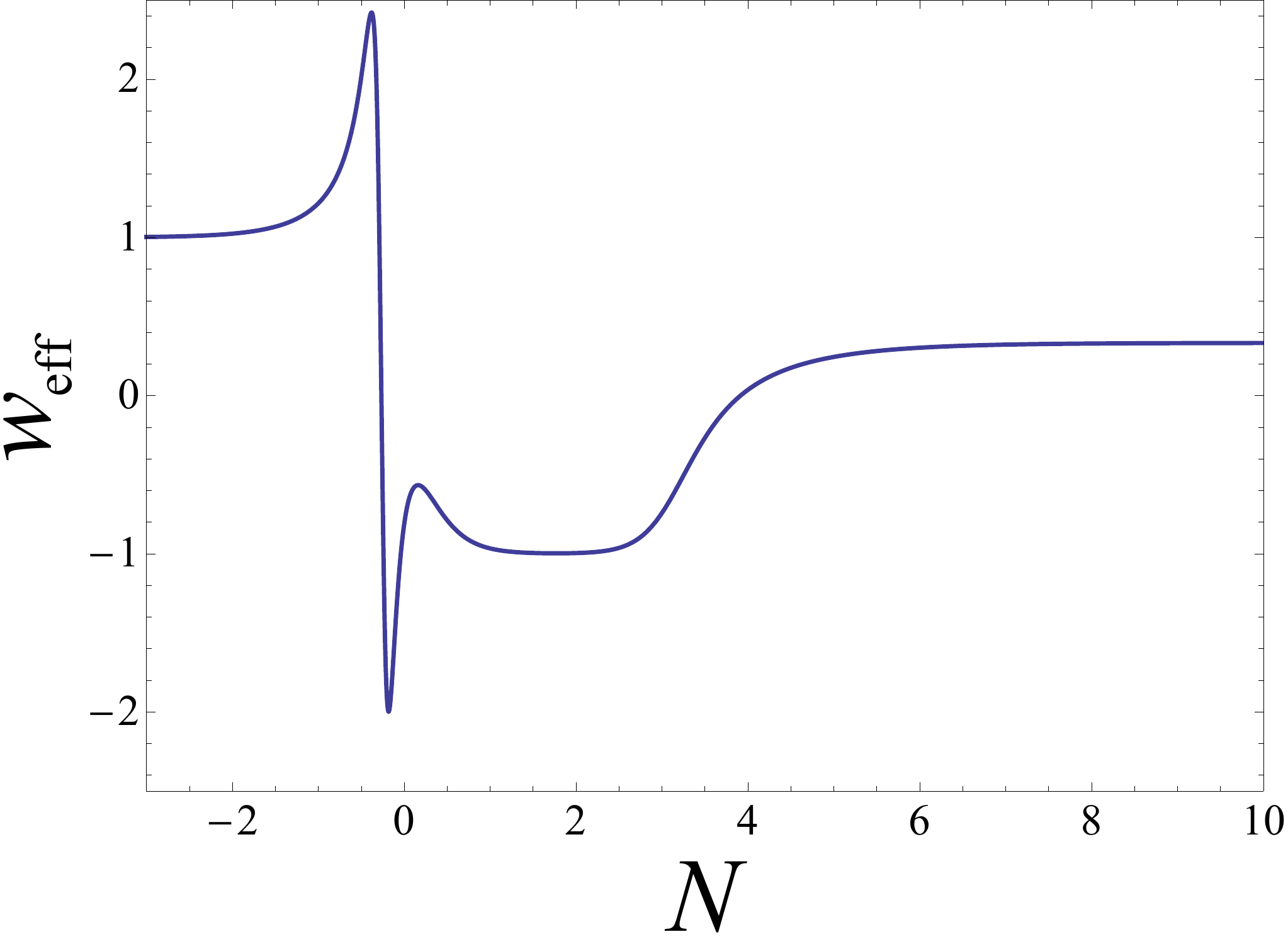}

\caption{Evolution of the effective equation of state ($w_{\rm eff}$) for interacting model I with potential $V=V_0 \sinh^{-\eta}(\lambda\phi)$. Here we have taken $w=\frac{1}{3}$, $\alpha=3$, $\beta=2\sqrt{3}$, $\lambda=1$, $\eta=2$.
} 
\label{fig:weff3_phn_cross}
\end{figure}

As in the case $\alpha=1$, critical points $A_{1\pm}$, $A_2$, $A_3$, $A_4$, $A_5$, $A_6$ will have each exactly two copies for two solutions $s_*=\pm \eta\lambda$.
Point $O$ always exists and its properties do not depend on the scalar field potential.
Point $A_{1+}$ is an unstable node whenever $\beta<-\frac{\sqrt{6}}{2}(3w+1)$, $\eta\lambda<\sqrt{6}$ and $\lambda<0$, otherwise it is a saddle, whereas point $A_{1-}$ behaves as an unstable node whenever $\beta>\frac{\sqrt{6}}{2}(3w+1)$, $\eta\lambda>-\sqrt{6}$ and $\lambda>0$, otherwise it behaves as a saddle. The regions of stability of points $A_2$, $A_4$ and $A_6$ in $(\beta,\lambda)$ parameter  space are given in Fig. \ref{region_sinh3} for $w=0$ and $\eta=1$.
Point $A_2$ corresponds to an unacclerated scaling solution with $w_{\rm eff}=w$. It is a late time attractor for some values of parameters $\beta$, $\lambda$. For example if we numerically choose $w=0$, $\beta=1$, $\eta=1$, $\lambda=3$ we obtain eigenvalues $E_1=-7$,  $E_2=-6$, $E_3=-0.75+1.56 i$, $E_3=-0.75-1.56 i$.
Point $A_3$ represents an accelerated universe whenever $\eta^2\lambda^2<2$. It is a stable node when $\eta^2\lambda^2<3(w+1)$, $\eta^2\lambda^2<\eta\lambda\beta+(9w+1)$ and $\eta>0$, otherwise it is a saddle.
Point $A_4$ corresponds to an accelerated solution. It is a late time accelerated attractor for some values of parameters $\beta$, $\lambda$ as outline in Fig. \ref{region_sinh3}. For example if we numerically choose $w=0$, $\beta=-2$, $\eta=1$, $\lambda=2$, we obtain eigenvalues $E_1=-3.5$, $E_2=-2.5$, $E_3=-16$, $E_4=-2$ and $w_{\rm eff}=-0.66$.
Due to complicated eigenvalues of point $A_5$, we consider only the physically interesting case $w=0$, for which it can be seen that this critical point is a saddle.
In a similar fashion, due to the complexity of critical point $A_6$, we will study its stability only for a pressure-less matter fluid.
It corresponds to an accelerated universe when $\frac{\eta\lambda}{\beta}>-\frac{2}{7}$. It is either a stable or a saddle node depending on the values of $\eta$, $\lambda$ and $\beta$ (see Fig. \ref{region_sinh3}).
Critical point $A_7$ corresponds to a late time attractor whenever $\eta<0$, it is a saddle whenever $\eta>0$.

To summarise, depending on the initial conditions, the Universe can evolve from a stiff matter solution $A_{1\pm}$ to either a scaling solution $A_2$, to an accelerated scalar field dominated solution $A_3$ (or $A_7$), to an accelerated attractor $A_4$, or to an accelerated attractor $A_6$ through a matter dominated solution $O$.
Thus, this model can describe a transition of the Universe from a matter dominated phase to a DE dominated phase. This phenomena can be clearly seen from Fig.~\ref{fig:weff_I_3}, which shows that after a long lasting period of matter domination ($w_{\rm eff}=0$), the Universe is then dominated by DE ($w_{\rm eff}=-1$).
On the other hand this model can also be used to describe an early inflationary cosmological phase with a crossing of the phantom barrier, as shown in Fig.~\ref{fig:weff3_phn_cross}.
This scenario is represented by trajectories starting from a stiff dominated solution, then evolving towards point $A_4$, which characterises the inflationary era, and finally ending in the late time attractor point $A_2$, where radiation domination is attained for $w = 1/3$.
As it is clear from Fig.~\ref{fig:weff3_phn_cross}, phantom crossing might be achieved before a period of standard inflation, whose duration depends on the initial conditions.
Moreover a graceful exit from inflation is automatically obtained since radiation domination is achieved right after the period of accelerated expansion without the need of a reheating process.
This model can thus be applied to both late time and early time phenomenology with distinguishing features that might provide useful observational signatures to look for in the astronomical data.


\subsubsection*{Example II: $ V=\frac{M^{4+n}}{\phi^n}$}


As discussed in the case $\alpha=1$, for this potential $s_*=0$.
All critical points except point $A_2$ exist and all of them are non-hyperbolic (see Table~\ref{tab:eigen_alg_A}).
Point $A_{1+}$ behaves as an unstable node whenever $\beta<-\frac{\sqrt{6}}{2}(3w+1)$ otherwise it is a saddle, whereas point $A_{1-}$ behaves as an unstable node whenever $\beta>\frac{\sqrt{6}}{2}(3w+1)$ otherwise it is a saddle.
Critical point $A_3$ coincides with $A_7$. This point is non-hyperbolic for this potential as $s_*=0$ and $g(0)=0$.
The stability of the center manifold of point $A_3$ is analyzed in the appendix A. As in the case of $\alpha=1$ it can be seen that this point corresponds to a late time attractor if $\Gamma(0)>1$ (i.e.~if $n>0$).
Linear analysis fails to determine the stability of the non-hyperbolic critical point $A_4$. However, due to its complicated expression we use numerical perturbation methods to check its stability, rather than doing it analytically with the centre manifold techniques. These numerical methods have been used in several cosmological models and found to be successful to determine the stability of complicated non-hyperbolic critical points \cite{Dutta:2016dnt,Dutta:2016bbs,Roy:2015cna}. Numerically we thus observe that perturbed trajectories around critical point $A_4$ do not asymptotically approach its coordinates, instead they are always attracted to the coordinates of point $A_7$ (or $A_3$). We can therefore conclude that critical point $A_4$ is not stable.
Critical point $A_5$ is not stable as its eigenvalue $E_1$ is positive whereas critical point $A_6$ behaves as a saddle as its eigenvalues $E_2$ and $E_3$ are of opposite sign.

From the above analysis, we see that, choosing a suitable set of model parameters, the Universe can undergo a transition from a matter dominated phase $O$ to a DE dominated phase $A_3$, meaning that in this case the model can successfully describe the late time behaviour of the Universe.

\section{Model II}\label{sec:model2}

This section deals with the analysis of model II of the dynamical system~(\ref{x_alg})-(\ref{s_alg}) where we assume the interacting energy density to take the simple form $\rho_{\rm int}=\gamma \phi\rho$, where $\gamma$ is a dimensionless parameter; see Table~\ref{models}. As before we can define the effective EoS parameter in this model to be
\begin{align}
w_{\rm eff}=w-(w-1)x^2-(1+w)y^2 \,.
\end{align}
Note that $w_{\rm eff}$ is independent of the interacting energy component since it does not depend on $z$.

\begin{table}
\begin{tabular}{cccccccc}
  \hline\hline
  Point ~~&~~ $x$~~~~ &~~~~~~ $y$~~~~~~ & ~~~~$z$~~~~&~~~~$s$~~~~&~~~~Existence~~~~&~~~~$w_{\rm eff}$ ~~~~&~~~~Acceleration\\
  \hline
  $B_{1\pm}$ & $\pm 1$ & 0 & 0&$s_*$&Always&$1$&No \\
  $B_2$ & $\sqrt{\frac{3}{2}} \frac{(1+w)}{s_*}$ &
  $\sqrt{\frac{3}{2}} \frac{\sqrt{(1+w)(1-w)}}{s_*}$ & $\frac{s_*^2-3(1+w)}{s_*^2}$&$s_*$ &Always&$w$&No\\
  $B_3$ & $\frac{s_*}{\sqrt{6}}$ & $\sqrt{1-\frac{s_*^2}{6}}$ & $0$ &$s_*$&$s_*^2 \leq 6$&$\frac{s_*^2}{3}-1$&$s_*^2<2$\\
  $B_4$ & 0 & 0 & 1 &$s$&Always&$w$&No\\
  $B_5$&$0$&$1$&$0$&$0$&Always&$-1$&Always\\
  \hline\hline
\end{tabular}
\caption{Critical points of Model II.}
\label{tab:alg_B}
\end{table}

\begin{table}
$\Xi_{\pm}=-\frac{3}{4}(1-w)\pm\frac{3}{4s_*}\sqrt{(1-w)(24(1+w)^2-s_*^2(7+9w))}$
\begin{tabular}{cccccc}
\hline\hline
Point &  $E_1$&$E_2$&$E_3$&$E_4$ & Stability \\
\hline
$B_{1\pm}$   &$3(1-w)$&$3(1-w)$&$3 \mp \sqrt{\frac{3}{2}}s_*$ &$\mp \sqrt{6}dg(s_*)$& Unstable node/saddle \\
$B_2$ & $0$&$\Xi_+$ &$\Xi_-$ &$-\frac{3(w+1)dg(s_*)}{s_*}$& Non-hyperbolic   \\
$B_3$ & $\frac{s_*^2-6}{2}$&$s_*^2-3(1+w)$&   $s_*^2-3(1+w)$&$-s_* dg(s_*)$&Stable node/saddle \\
$B_4$ & $0$&$0$&$\frac{3}{2}(w-1)$&$\frac{3}{2}(w+1)$ & Saddle \\[1.5ex]
&&&&& Stable if $g(0)>0$\\
$B_5$ & $-3(w+1)$&$-3(w+1)$& $\frac{1}{2}\left(-\sqrt{12 g(0)+9}-3\right)$& $\frac{1}{2}\left(\sqrt{12 g(0)+9}-3\right)$&Saddle if $g(0)<0$\\
&&&&& See App. B if $g(0)=0$\\
\hline\hline
\end{tabular}
\caption{Stability of critical points of Model II.}
\label{tab:eigen_alg_B}
\end{table}

The existence and cosmological properties of the critical points for this model are given in Table~\ref{tab:alg_B}.
There are up to six critical points, and all critical points are independent of the parameter $\gamma$.
The corresponding eigenvalues of the critical points, determining their stability properties, are given in Table~\ref{tab:eigen_alg_B}. Critical points $B_{1\pm}$, $B_2$ and $B_3$ depend on the particular potential considered.
Point $B_2$ exists for any values of $s_*$ and $w$ in contrast to the canonical case without interaction, whereas point $B_3$ exists only when $s_*^2\leq6$.
Critical point $B_4$ exists for any arbitrary potentials.
On the other hand critical point $B_5$ corresponds to the case where the scalar field potential is effectively constant as the $\phi$-derivative of the potential vanishes. Nevertheless its stability depends on the concrete form of the potential.
Note that critical point $B_3$ reduces to $B_5$ when $s_*=0$.

Critical points $B_{1\pm}$ correspond to a scalar field kinetic energy dominated solutions with stiff fluid effective EoS ($w_{\rm eff}=1$).
Critical point $B_{1+}$ is an unstable node when $s_*<\sqrt{6}$ and $dg(s_*)>0$, whereas point $B_{1-}$ is an unstable node when $s_{*}>-\sqrt{6}$ and $dg(s_*)<0$, otherwise they are both saddles.
Critical point $B_2$ represents a solution dominated by the scalar field energy density and the interacting energy density, with the Universe behaving as if it was matter dominated ($w_{\rm eff}=w$).
This point is non-hyperbolic due to at least one vanishing eigenvalue.
Therefore, one cannot determine the stability of the point using linear stability theory.
Due to the complicated expressions of the eigenvalues of critical point $B_2$, we will not apply center manifold theory here.
Instead we will study its stability numerically, postponing the analysis once a specific potential has been selected.
Critical point $B_3$ corresponds to a scalar field dominated universe.
This point describes an accelerated universe whenever $s_*^2<2$.
It is a stable node when $s_*^2<3(1+w)$ and $s_*dg(s_*)>0$, otherwise it is a saddle.
Point $B_4$ stands for a solution dominated by the interacting energy density with vanishing scalar field and matter energy densities, where the universe behaves as if it was matter dominated ($w_{\rm eff}=w$).
This point is non-hyperbolic but it behaves as a saddle since the non-vanishing eigenvalues are opposite in sign.
Critical point $B_5$ corresponds to an accelerated scalar field dominated solution. It is stable whenever $g(0)>0$, it is a saddle whenever $g(0)<0$ and it is non-hyperbolic in nature for $g(0)=0$. For this latter case the full analysis of the stability of this point using center manifold theory is given in appendix B. Again this point corresponds to a late time scalar field dominated attractor if $\Gamma(0)>1$.

From the above analysis, we see that the dynamics of this model is similar to the case of canonical scalar field without interaction. In this case, the Universe evolves from stiff matter solutions $B_{1\pm}$ and evolves toward a scaling solution $B_2$ or an accelerated scalar field dominated solution $B_3$ (or $B_5$). Note that points $B_2$ and $B_3$ cannot be late time attractors simultaneously as $B_2$ is not stable when $s_*^2<3(1+w)$.
Interestingly, the matter dominated phase, usually described by the origin $O$, is in this case replaced by point $B_4$, corresponding to an interaction dominated phase with $w_{\rm eff}=w$.
In order to deeper understand the role of the scalar field potential on the cosmological dynamics of this model and to analyze in detail the stability of non-hyperbolic critical points, in what follows we consider two distinct concrete potentials in analogy to what done for model~I.

\begin{figure}
\centering
\subfigure[]{%
\includegraphics[width=8cm,height=6cm]{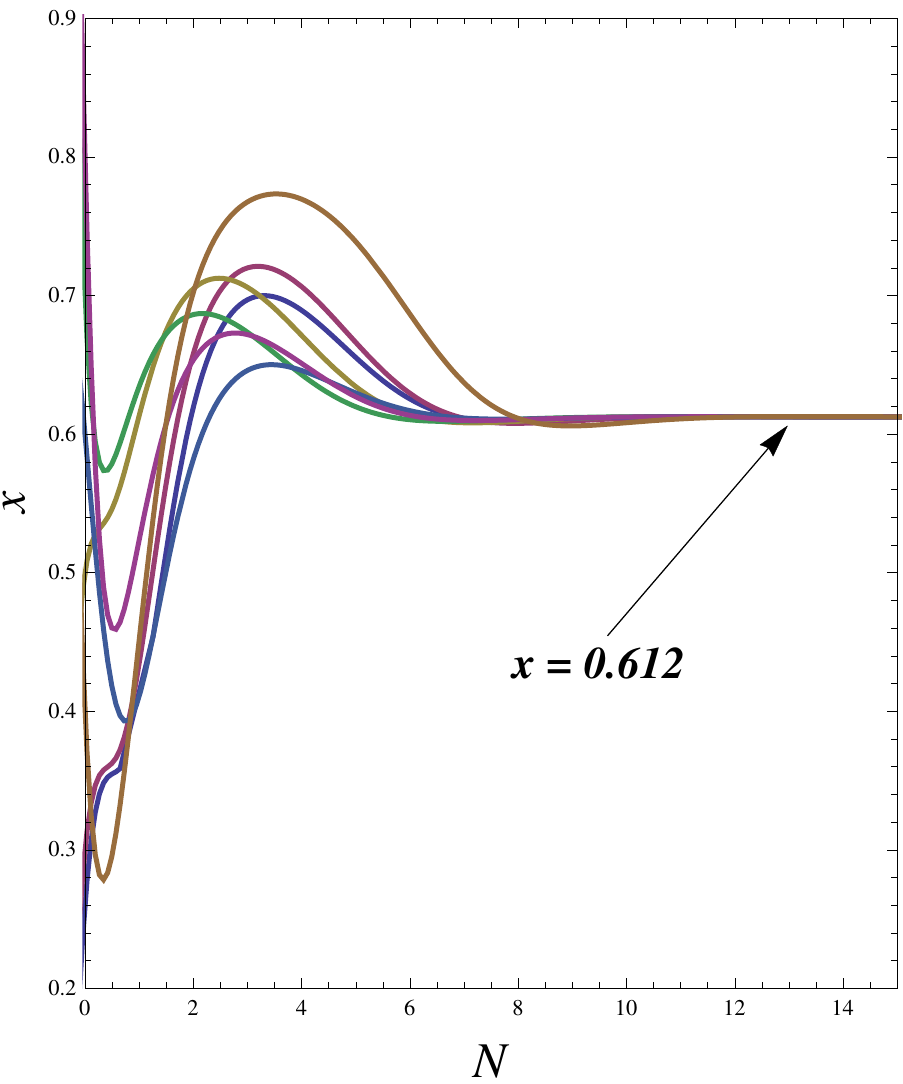}\label{fig1}}
\qquad
\subfigure[]{%
\includegraphics[width=8cm,height=6cm]{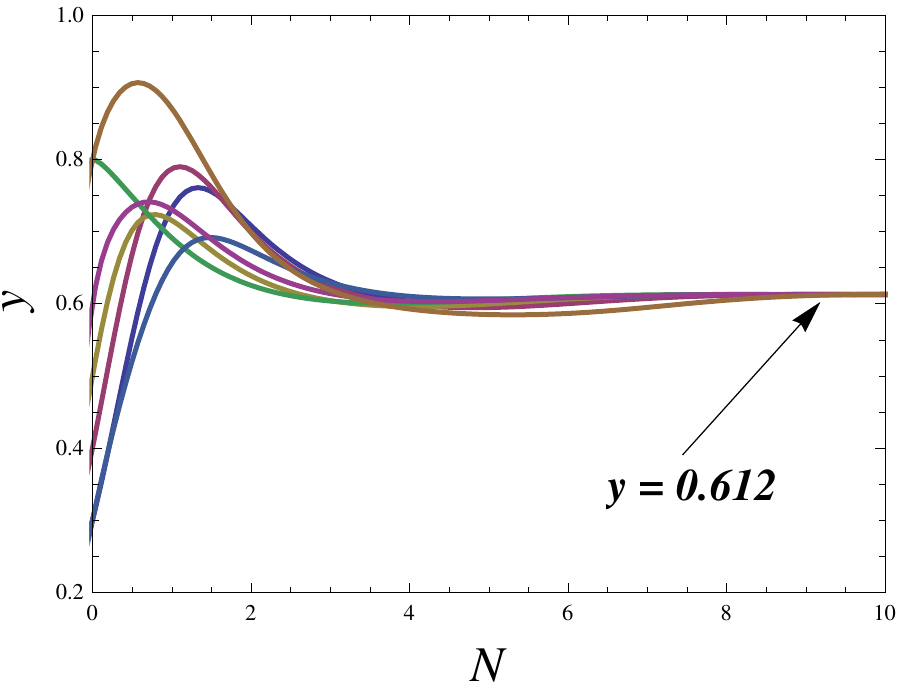}\label{fig2}}
\qquad
\subfigure[]{%
\includegraphics[width=8cm,height=6cm]{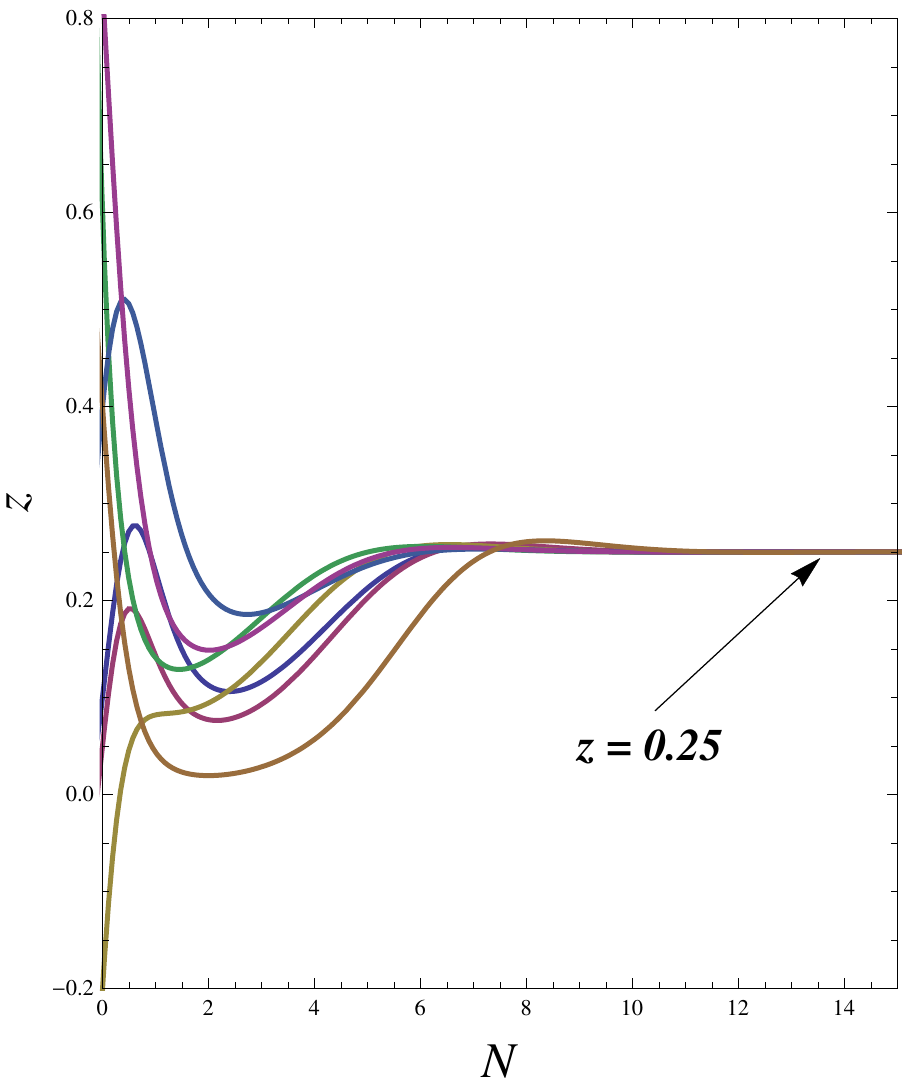}\label{fig3}}
\qquad
\subfigure[]{%
\includegraphics[width=8cm,height=6cm]{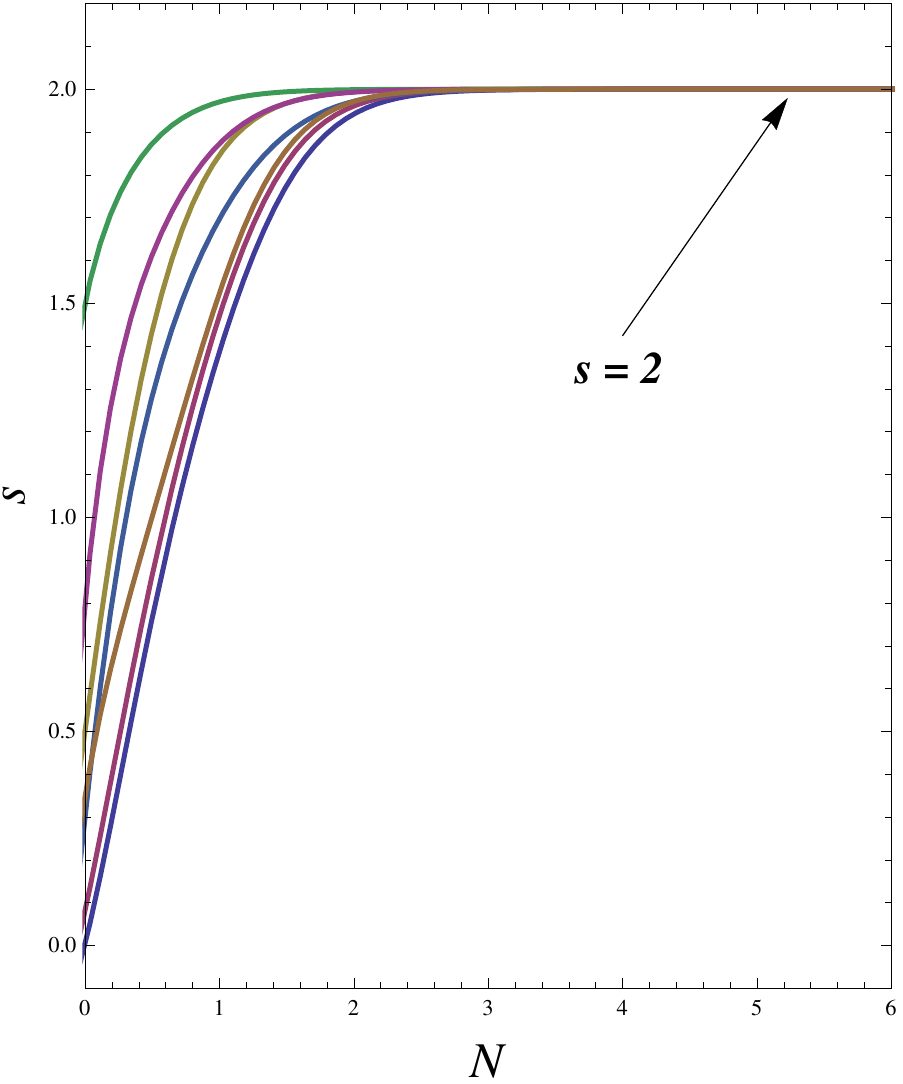}\label{fig4}}
\caption{(a).~Time evolution of trajectories projected on $x$-axis approaching point $B_2$; (b).~Time evolution of of trajectories projected on $y$-axis approaching point $B_2$; (c).~Time evolution of trajectories projected on $z$-axis approaching point $B_2$; (d).~Time evolution of trajectories projected on $s$-axis approaching point $B_2$. Here we have considered the potential $V=V_0\,\sinh^{-\eta}(\lambda\phi)$ (i.e.,~$g(s)=\frac{1}{\eta}-\frac{\eta\lambda^2}{s^2}$, $s_*=\pm \eta\lambda$, $dg(s_*)=\frac{2\eta\lambda^2}{s_*^3}$) with $\gamma=1$, $\eta=2$, $\lambda=1$ and $w=0$.}
\end{figure}\label{pert_model_II_B}

\subsubsection*{Example I: $V=V_0\sinh^{-\eta}(\lambda\phi)$}

\begin{figure}
\centering
\includegraphics[width=8cm,height=6cm]{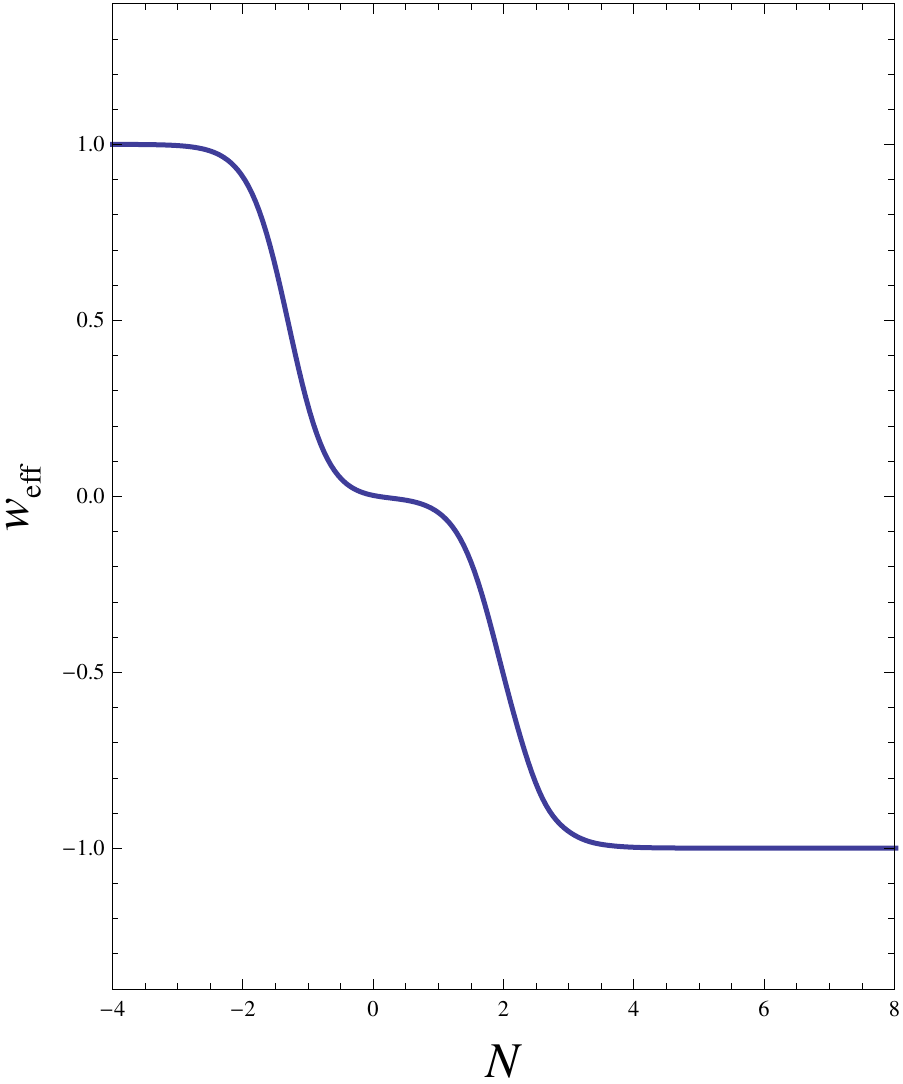}
\caption{Plot of $w_{\rm eff}$ versus $N$ of model II with potential $ V=V_0\sinh^{-\eta}(\lambda\phi)$ showing a transition from interaction dominated epoch (point $B_4$) with $w_{\rm eff}=0$ to a DE dominated epoch (point $B_5$). Here we have taken $w=0$, $\gamma=1$, $\lambda=4$, $\eta=-0.5$.}
\label{fig:weff_2}
\end{figure}

In this case critical points $B_{1\pm}$, $B_2$, $B_3$ will have each exactly two copies for two solutions $s_*=\pm \eta\lambda$, while point $B_4$ exist for any arbitrary potentials. Point $B_5$ does not exist for this potential.
Critical point $B_{1+}$ is an unstable node when $\eta\lambda<\sqrt{6}$ and $\lambda>0$, whereas point $B_{1-}$ is an unstable node when $\eta\lambda>-\sqrt{6}$ and $\lambda<0$, otherwise they are saddles.
The stability of critical point $B_2$ can now be determined in detail using numerical perturbation technique.
In Figs.~\ref{fig1}-\ref{fig4}, we have numerically plotted the projections of the phase space on the $x$, $y$, $z$, and $s$ axes separately.
We observe that depending on the values of parameters $\lambda$ and $\eta$, the perturbed solutions will asymptotically approach the point $B_2$ as $N\rightarrow \infty$.
Critical point $B_3$ corresponds to an accelerated universe whenever $\eta^2\lambda^2<2$. It is a stable node when $\eta^2\lambda^2<3(1+w)$ and $\eta>0$, otherwise it is a saddle.
Critical point $B_4$ behaves as a saddle. Critical point $B_5$ is stable when $\eta<0$, it is a saddle whenever $\eta>0$.
Thus, we find that for some range values of the parameters, the observed late time behaviour of our Universe can be successfully described by this model, as shown for example by Fig.~\ref{fig:weff_2} where the values $w=0$, $\gamma=1$, $\lambda=4$, $\eta=-0.5$ have been chosen.

\subsubsection*{Example II: $ V=\frac{M^{4+n}}{\phi^n}$}

As discussed earlier for this potential $s_*=0$, and so critical point $B_2$ does not exist. In this case, point $B_3$ coincides with point $B_5$. All remaining critical points are non-hyperbolic points.
Critical points $B_{1\pm}$ behaves as an unstable node, whereas critical point $B_4$ behaves as a saddle.
The stability of point $B_3$ (or $B_5$) is determined using center manifold theory (since the eigenvalues $E_2$, $E_3$ vanish as $g(0)=0$).
The full analysis is shown in appendix B and leads to the result that this point corresponds to a late time attractor if $\Gamma(0)>1$ (i.e.~if $n>0$). Hence, we see that the Universe can evolve from a stiff matter dominated phase towards a DE dominated phase through a matter phase dominated by the interaction between DE and DM for a wide range of initial conditions.
The evolution is in this case similar to the one found in the previous example (see Fig.~\ref{fig:weff_2}) and roughly coincides with the one described by $\Lambda$CDM, although during the matter phase the interaction between DE and DM is present and thus differences might arise at the level of perturbations for example.


\section{Conclusion}\label{sec:conc}

In the present paper, we have employed dynamical systems techniques to study the background cosmological evolution of two interacting DE models derived from a newly proposed variational method.
This type of interacting theories, named Scalar-Fluid theories, is constructed at the Lagrangian level and thus is well motivated by an underlining theoretical framework, even though not a fundamental one.
In analogy with the investigation of \cite{Boehmer:2015kta}, in our analysis we have considered two different algebraic coupling functions defining the DM-DE interaction within the Scalar-Fluid action; see Table~\ref{models}.
In fact the main scope of our work has been to extend the dynamical analysis of \cite{Boehmer:2015kta}, performed only for a scalar field exponential potential, to a broader class of potentials.
This in general leads to higher dimensional dynamical systems than the three-dimensional ones obtained with an exponential potential.
We found that these extended autonomous systems contain more critical points than the cases studied in \cite{Boehmer:2015kta}.
We also found that there are some critical points which depend on the specific form of the potential for their existence (e.g., $A_{1\pm}- A_6$, $B_{1\pm} - B_3$), or their stability (e.g., $A_7$ and $B_5$), while other critical points are independent of the specific form of the scalar field potential (e.g., points $O$, $B_4$).
For this type of extension, we also obtained some non-hyperbolic critical points such as points $A_7$ (see Table \ref{tab:eigen_alg_A}), $B_2$ and $B_5$ (see Table \ref{tab:eigen_alg_B}).
For these points we have used center manifold theory or numerical methods of perturbed trajectories to determine their stability.
In order to better understand the cosmological dynamics, we also have considered two concrete potentials as examples: the hyperbolic potential $V=V_0\sinh^{-\eta}(\lambda\phi)$ and the power law potential $V=\frac{M^{4+n}}{\phi^n}$.

In model I, where the DE-DM coupling term in the Scalar-Fluid Lagrangian is chosen to be $\rho_{\rm int} = \gamma\,\rho^\alpha\exp(-\beta\phi)$, we obtained physically interesting solutions like a standard matter dominated solution (point $O$), late time accelerated scalar field dominated solutions (points $A_3$, $A_7$) and also a late time accelerated scaling solution (point $A_6$) which can possibly alleviate the coincidence problem (see Sec. \ref{alpha1}).
The behaviour of this last critical point is similar to the one generally arising in standard interacting DE models, where accelerating scaling solutions are usually attained.
Model I can also be employed to describe an inflationary era with interesting features, for example crossing of the phantom barrier (see \ref{subsec:model_I_3}).
This shows that the model can also be applied to explain the observed phenomenology at early times, with possible distinguishing signatures that might be present in the astronomical data.
Finally in model I (with $\alpha=3$), we also obtain multiple late time attractors for some choices of the parameters.
These kinds of situations are usually of great theoretical and mathematical interest, for example regarding the choice of initial conditions and bifurcation theory.

The background cosmological dynamics of model II, where the $\rho_{\rm int} = \gamma\phi\rho$ coupling is considered, is similar to the case of an uncoupled standard scalar field model, except that the matter dominated solution is replaced by an interacting dominated solution between DE and DM (point $B_4$).
This scenario well reproduces the observed dynamics of the universe and, although undistinguishable at the background level from the standard $\Lambda$CDM evolution, might produce differences at the perturbation level.
Cosmological perturbation analysis and comparison against astronomical observations constitute the next natural step in the investigations of these interacting models, but they lie beyond the scope of our present study, and will be left for future works.

To summarise, the analysis presented in this paper reveals that the results of \cite{Boehmer:2015kta} obtained with an exponential potential only, can be equally derived for other scalar field potentials, similarly e.g.~to the case of quintessence \cite{Fang:2008fw} and $k$-essence \cite{Dutta:2016bbs}.
On a physical ground we note in fact that in both Model I and Model II a matter to DE transition can be achieved at late times.
This situation well describes the observed background behaviour of our Universe and could in principle produce detectable discrepancies from $\Lambda$CDM at the linear (or non-linear) perturbation levels.

\appendix

\section*{Appendix: Center Manifold Theory (CMT)}
\subsection*{Mathematical Background}

In this section we briefly briefly review the mathematical background of CMT. The detailed mathematical background with examples is given in \cite{Wiggins,Perko}, while similar applications to cosmology can be found in \cite{Boehmer:2011tp,TamaniniPhDthesis}.
Linear stability theory fails to determine the stability of a critical point whose stability matrix contains at least one vanishing eigenvalue.
If at least (the real part of) one of the non vanishing eigenvalues is positive then the critical point is unstable, but if all non vanishing eigenvalues are negative then stability is not guaranteed.
In these cases CMT can be applied.
The main objective of CMT is to investigate the dynamics on the dimensionally reduced space identified by the eigenvectors corresponding to the vanishing eigenvalues.
This reduced space is called the center manifold and its existence is always guaranteed.

Without any loss of generality we assume that the critical point is the origin (this can always be achieved by translating the origin to the critical point). Any arbitrary non-linear system of differential equations with only negative and vanishing eigenvalues in the stability matrix can be written as
\begin{align}
u'&=A u+f(u,v)\label{cmteqn1} \\
v'&=B v+g(u,v)\label{cmteqn2}
\end{align}
where $(u,v)\in \mathbb{R}^c \times \mathbb{R}^s$ with $f$ and $g$ satisfying
\begin{align*}
f(0,0)=0,\quad Df(0,0)=0\\
g(0,0)=0,\quad Dg(0,0)=0
\end{align*}
Here $A$ is a $c\times c$ matrix with eigenvalues having zero real part ($c$ is the number of vanishing eigenvalues), $B$ is a $s\times s$ matrix with eigenvalues having negative real part ($s$ is the number of non vanishing eigenvalues) and $Df$ denotes the Jacobian matrix of $f$.
The center manifold is characterized by a function $h: \mathbb{R}^c\rightarrow \mathbb{R}^s$ and is defined as:
\begin{align}
W^c(0)=\left\lbrace (u,v)\in \mathbb{R}^c\times \mathbb{R}^s: v=h(u), \Vert u \Vert < \delta, h(0)=0, D h(0)=0\right\rbrace \,,
\end{align}
for a sufficiently small $\delta$. $h$ is at least of class $C^2$ and $\Vert\,\cdot\, \Vert$ denotes the Euclidean norm.
The dynamics of the system (\ref{cmteqn1})-(\ref{cmteqn2}) restricted to the center manifold $W^c(0)$ is determine by the equation
 \begin{align}\label{cmtreduce}
 u'=A u+f(u,h(u))
 \end{align}
for a sufficiently small $u \in \mathbb{R}^c$. The stability/instability of the system (\ref{cmtreduce}) implies the stability/instability of the original system (\ref{cmteqn1})-(\ref{cmteqn2}).
The problem now is how to determine $h$. It can be proven that $h$ must satisfy the following quasilinear partial differential equation \cite{Wiggins,Perko}
 \begin{align}\label{quasi}
 \mathcal{N}h(u)\equiv Dh(u)\left(A u+f(u,h(u))-B h(u)-g(u,h(u))\right)=0
\end{align}
In general it is often impossible to solve $h$ from Eq.~(\ref{quasi}) analytically. Fortunately, the solution $h$ of (\ref{quasi}) can be approximate by a power series expansion valid up to a desired degree of accuracy (see \cite{Wiggins,Perko}).
In order to explain how this works, we shall now apply this method to determine the stability of point $A_7$ for model~I and point $B_6$ of model~II.

\subsection*{Appendix A: Center manifold dynamics for point $A_7$ of model I}

In this appendix, we apply center manifold theory to study the stability of point $A_7=(0,1,0,0)$ when $g(0)=0$ for general values of $\alpha$. As mentioned in the main text this point is a saddle whenever $\alpha<0$. Firstly, we translate the point $(0,1,0,0)$ to the origin by using the transformation $x\rightarrow x$, $y\rightarrow y+1$, $z\rightarrow z$, $s\rightarrow s$. Eqs.~(\ref{x_alg})--(\ref{s_alg}) then become
\begin{align}
 x' &= -\frac{1}{2} \left(3x \left((w+1) (y+1)^2+w z-w+1\right) + 3 (w-1) x^3-\sqrt{6} s  (y+1)^2\right) + \frac{3}{2} x z \left(\alpha(w+1)-1\right)+\sqrt{\frac{3}{2}}\beta z, \\
  y' &=-\frac{1}{2} (y+1) \left( 3 (w-1) x^2 + 3 \left((w+1) (y+1)^2+w z - w-1\right)+\sqrt{6} s  x\right) + (y+1) \frac{3}{2} z \left(\alpha(w+1)-1\right), \\
  z' &= 3 z \left(\alpha(w+1)-1\right) (z-1)-2 \sqrt{\frac{3}{2}}\beta z x-3 z \left((w-1) x^2 + (w+1) (y+1)^2 + w (z-1)\right),\\
  s'&=-\sqrt{6}\, x\, g(s),
\end{align}
Using the eigenvectors of the stability matrix of $A_7$ in the transformed system to form a new basis, we now introduce a new set of variables given by
\[\left(\begin{array}{c}
X\\
Y\\
Z\\
S \end{array} \right)=\left(\begin{array}{cccc}
1                 & 0 &  \frac{1}{6}\frac{\beta \sqrt{6}}{\alpha(w+1)-1}  &   -\frac{1}{\sqrt{6}}\\
0                 & 1 &  -\frac{1}{2}                                     & 0\\
0                 & 0 & 1                                                &  0\\
0                 & 0 & 0                                                &  1 \end{array} \right) \left(\begin{array}{c}
x\\
y\\
z\\
s \end{array} \right)\]
In terms of these new set of variables, the system of equations can now be written as
\[\left(\begin{array}{c}
X'\\
Y'\\
Z'\\
S' \end{array} \right)=\left(\begin{array}{cccc}
-3  & 0 & 0  &  0\\
0  & -3(w+1) & 0 & 0\\
0  & 0 &  -3\alpha (w+1) &  0\\
0  & 0 & 0  & 0 \end{array} \right) \left(\begin{array}{c}
X\\
Y\\
Z\\
S \end{array} \right)+\left(\begin{array}{c}
g_1\\
g_2\\
g_3\\
f \end{array} \right)\]
where $f,\,g_1,\,g_2,\,g_3$ are polynomials of degree greater than 2 in $(X,\,Y,\,Z,\,S)$ with
\begin{equation}
f(X,Y,Z,S)=-{\frac {\left( \Gamma(S)-1 \right) {S}^{2} \left( \sqrt {6}\alpha\,X (w+1)+\alpha\, S (w+1)-\sqrt {6} X-\beta\,Z-S \right) }{\alpha\,(w+1)-1}},
\end{equation}
whereas $g_1$, $g_2$ and $g_3$ are not shown due to their lengths.
Note that the dynamical system is now in the form \eqref{cmteqn1}--\eqref{cmteqn2}.
At this point the coordinates which correspond to non-zero eigenvalues $(X,Y,Z)$ can be approximated in terms of $S$ by functions
\begin{equation}
h_1(S)=a_2 S^2+a_3 S^3+\mathcal{O}(S^4),
\end{equation}
\begin{equation}
h_2(S)=b_2 S^2+b_3 S^3+\mathcal{O}(S^4),
\end{equation}
\begin{equation}
h_3(S)=c_2 S^2+c_3 S^3+\mathcal{O}(S^4),
\end{equation}
respectively. Thus the quasilinear partial differential equation which the functions \[\mathbf{h}=\left(\begin{array}{c}
h_1\\
h_2\\
h_3 \end{array} \right)\] have to satisfy is given by
\begin{equation}\label{quasiI}
D \mathbf{h(S)}\left[A S+\mathbf{F}(S,\mathbf{h}(S))\right]-B \mathbf{h}(S)-\mathbf{g}(S,\mathbf{h}(S))=\mathbf{0}
\end{equation}
Here, \[\mathbf{g}=\left(\begin{array}{c}
g_1\\
g_2\\
g_3 \end{array} \right),~~~~~ \mathbf{F}=g, ~~~~~B= \left(\begin{array}{ccc}
 -3 & 0 & 0\\
 0 & -3(w+1) &  0\\
 0 & 0  &  -3\alpha (w+1) \end{array} \right), \quad A=0 \,. \]
In order to solve the Eq.~(\ref{quasi}), we substitute $A$, $\textbf{h}$, $\mathbf{F}$, $B$, $\mathbf{g}$ into it and compare equal powers of $S$ in order to obtain the series that approximates $\mathbf{h}(S)$. Thus on comparing powers of $S$ from both sides of Eq.~(\ref{quasi}) we obtain the constants $a_2$, $a_3$, $b_2$, $b_3$, $c_2$, $c_3$ where
\begin{align}
a_2=0,~ a_3=\frac{\Gamma(0)-1}{3\sqrt{6}}, ~~ b_2=-\frac{1}{12},~~ b_3=0,~~ c_2=0,~~ c_3=0.
\end{align}
Now, the dynamics of the reduced system is determined by the equation
\begin{align}
S'=A\,S+\mathbf{F}(S,\mathbf{h}(S)) \,,
\end{align}
namely
\begin{align}
S'&=-\left(\Gamma(0)-1\right) S^3+\mathcal{O}(S^4) \,.
\end{align}
From this last equation we can immediately conclude that point $A_7$ corresponds to a late time attractor if $\Gamma(0)>1$ and $\alpha>0$.

\subsection*{Appendix B: Center manifold dynamics for point $B_5$ of model II}

In this appendix we apply center manifold theory to study the stability of point $B_5$ when $g(0)=0$. As in App.~A, we first translate the point $(0,1,0,0)$ to the origin by using the transformation $x\rightarrow x$, $y\rightarrow y+1$, $z\rightarrow z$, $s\rightarrow s$. Then Eqs.~(\ref{x_alg})-(\ref{s_alg}) become
\begin{align}
 x' &= -\frac{1}{2} \left(3x \left((w+1) (y+1)^2+w z-w+1\right) + 3 (w-1) x^3-\sqrt{6} s  (y+1)^2\right) + \frac{3}{2} x w z +\sqrt{\frac{3}{2}}\gamma (1-x^2-(y+1)^2-z), \\
  y' &=-\frac{1}{2} (y+1) \left( 3 (w-1) x^2 + 3 \left((w+1) (y+1)^2+w z - w-1\right)+\sqrt{6} s  x\right) + \frac{3}{2}  (y+1)\, z \left(\alpha(w+1)-1\right), \\
  z' &= 3 w z (z-1)-2 \sqrt{\frac{3}{2}}\gamma (1-x^2-(y+1)^2-z) x-3 z \left((w-1) x^2 + (w+1) (y+1)^2 + w (z-1)\right),\\
  s'&=-\sqrt{6}\, x\, g(s),
\end{align}
Using the eigenvectors of the Jacobian matrix of point $B_5$ in the transformed system as a new basis, we now introduce a new set of variables given by
\[\left(\begin{array}{c}
X\\
Y\\
Z\\
S \end{array} \right)=\left(\begin{array}{cccc}
1                 & \frac{1}{3}\frac{\gamma\sqrt{6}}{w} & \frac{1}{6}\frac{\gamma \sqrt{6}}{w}     &   -\frac{1}{\sqrt{6}}\\
0                 & 0                                    & 1                                        & 0\\
0                 & 1                                    & 0                                        &  0\\
0                 & 0                                    & 0                                       &  1 \end{array} \right) \left(\begin{array}{c}
x\\
y\\
z\\
s \end{array} \right)\]
In terms of these new set of variables the system of equations can now be written as
\[\left(\begin{array}{c}
X'\\
Y'\\
Z'\\
S' \end{array} \right)=\left(\begin{array}{cccc}
-3  & 0 & 0  &  0\\
0  & -3(w+1) & 0 & 0\\
0  & 0 & -3(w+1) &  0\\
0  & 0 & 0  & 0 \end{array} \right) \left(\begin{array}{c}
X\\
Y\\
Z\\
S \end{array} \right)+\left(\begin{array}{c}
g_1\\
g_2\\
g_3\\
f \end{array} \right)\]
where $f,\,g_1,\,g_2,\,g_3$ are polynomials of degree greater than 2 in $(X,\,Y,\,Z,\,S)$ with
\begin{equation}
f(X,Y,Z,S)=\frac{S^2 \left(\Gamma(S)-1\right)}{w}(\sqrt{6} w X+2\gamma Z-\gamma Y-S w),
\end{equation}
whereas $g_1$, $g_2$ and $g_3$ are again not shown due to their lengths.
Again, following the steps outlined in App.~A, by comparing powers of $S$ from both sides of Eq.~(\ref{quasi}) we obtain the constants $a_2$, $a_3$, $b_2$, $b_3$, $c_2$, $c_3$ where
\begin{align}
a_2=-\frac{1}{36}\frac{\gamma\sqrt{6}}{w},~ a_3=\frac{\Gamma(0)-1}{3\sqrt{6}}, ~~ b_2=0,~~ b_3=0,~~ c_2=-\frac{1}{12},~~ c_3=0.
\end{align}
The dynamics of the reduced system on the center manifold is thus determined by the equation
\begin{align}
S'&=-\left(\Gamma(0)-1\right) S^3+\mathcal{O}(S^4) \,.
\end{align}
We can then conclude that point $B_5$ corresponds to a late time attractor if $\Gamma(0)>1$, otherwise it is unstable (saddle).

\end{document}